\documentclass[aps,prl,twocolumn,showpacs,superscript address]{revtex4-1}
\usepackage{amsmath}
\usepackage{upgreek}
\usepackage{graphicx}
\usepackage[euler]{textgreek}
\usepackage[export]{adjustbox}
\usepackage{float}
\usepackage{physics}
\usepackage{mathrsfs}
\usepackage{bm}
\usepackage{caption}
\usepackage{textcomp}
\usepackage{natbib}
\usepackage{filecontents}
\usepackage{hyperref}

\graphicspath{ {/} }
\begin{document}
\captionsetup[figure]{labelsep=period,name={FIG.}, format={plain},justification={raggedright}}
\title{Ultrafast broadband optical spectroscopy for quantifying subpicometric coherent atomic displacements in WTe\textsubscript{2}}

\author{Davide Soranzio}
\affiliation{Dipartimento di Fisica, Universit\`a degli Studi di Trieste, 34127 Trieste, Italy}
\author{Maria Peressi}
\affiliation{Dipartimento di Fisica, Universit\`a degli Studi di Trieste, 34127 Trieste, Italy}
\author{Robert J. Cava}
\affiliation{Department of Chemistry, Princeton University, Princeton, New Jersey 08544, USA}
\author{Fulvio Parmigiani}
\affiliation{Dipartimento di Fisica, Universit\`a degli Studi di Trieste, 34127 Trieste, Italy}
\affiliation{International Faculty, University of Cologne, Albertus-Magnus-Platz, 50923 Cologne, Germany}
\affiliation{Elettra-Sincrotrone Trieste S.C.p.A., 34149 Basovizza, Italy}
\author{Federico Cilento}
\email[e-mail:]{federico.cilento@elettra.eu}
\affiliation{Elettra-Sincrotrone Trieste S.C.p.A., 34149 Basovizza, Italy}

\date{\today}

\begin{abstract}

Here we show how time-resolved broadband optical spectroscopy can be used to quantify, with femtometer resolution, the oscillation amplitudes of coherent phonons through a displacive model without free tuning parameters, except an overall scaling factor determined by comparison between experimental data and density functional theory calculations.
WTe\textsubscript{2} is used to benchmark this approach. In this semimetal, the response is anisotropic and provides the spectral fingerprints of two A\textsubscript{1} optical phonons at \protect\begin{math}\sim\protect\end{math}8 cm\textsuperscript{-1} and \protect\begin{math}\sim\protect\end{math}80 cm\textsuperscript{-1}. In principle, this methodology can be extended to any material in which an ultrafast excitation triggers coherent lattice modes modulating the high-energy optical properties. \end{abstract}

\maketitle

Several fundamental properties of materials, including the electrical and thermal conductivities, are influenced by lattice vibrations \cite{Ashcroft}. Recently, the possibility to control such behaviors by resonant coupling of ultrashort light pulses to specific lattice modes has been proved \cite{Rini2007}\cite{Nicoletti}. Conversely, coherent phonon spectroscopy has emerged as a powerful method to directly observe, in the time domain, coherent lattice vibrations \cite{Dekorsy2000}\cite{Ishioka2010}\cite{Kuznetsov1996}.
Despite the many studies, coherent phonons have been exploited primarily to characterize ground state properties \cite{Ishioka2006}\cite{Melnikov}\cite{Kim2015}\cite{Ishioka2016}, rather than controlling the material properties. This observation motivates the quest to learn how specific lattice vibrations affect the electronic behavior in the vicinity of the Fermi energy.
Structural dynamics experiments, such as time-resolved x-ray diffraction \cite{Sokolowski} and time-resolved electron diffraction \cite{ued}, are currently used to measure the amplitude of synchronized collective excitations of the atoms in a solid, i.e. coherent phonon modes. 
Yet, subpicometer displacements in complex materials are very challenging to be resolved.

In time-resolved reflectivity experiments, such modes appear as ultrafast oscillations of the probe signal \cite{Cheng}\cite{coherentphonons}. Their amplitude has been related to the atomic shifts for single-element materials  \cite{decamp}\cite{Katsuki2013} by  using single-frequency measurements. In materials with complex unit cells, a larger number of phonon modes is present and a method that avoids correlations in the estimation of individual amplitudes is required.
Here, we report on a novel approach to estimate the non-equilibrium atomic displacements of coherent optical phonon modes by detecting the modulation induced in a broadband reflectivity probe signal, which is central to obtain reliable signatures of the modes. Density functional theory (DFT) calculations are applied to a displacive model to simulate the time-resolved optical reflectivity signals and to estimate the coherent phonon amplitudes in the tens of femtometers regime.

At present, our method is applied to the orthorhombic semimetallic (Td) phase of tungsten ditelluride (WTe\textsubscript{2}), a transition metal dichalcogenide that has recently gained interest for showing an unusually high and non-saturating magnetoresistance \cite{magnetoWTe2} along with a possible type-II Weyl semimetal character \cite{weyl2}.
WTe\textsubscript{2} has a layered structure belonging to the space group Pmn2\textsubscript{1} \cite{cryst}
(see Fig. S1 in \cite{SI}) with in-plane covalent bonding and principally van der Waals interactions holding together the individual layers.
Furthermore, it has been shown that its electronic, optical and topological properties are influenced by strain forces \cite{weyl2}\cite{straintmdc}, which can be induced through non-equilibrium perturbations \cite{Sie2019} allowing an ultrafast control of the functionalities of this material.
 
Expressly, we focus on two A\textsubscript{1} coherent optical phonon modes at \begin{math}\sim\end{math}8 cm\textsuperscript{-1} and \begin{math}\sim\end{math}80 cm\textsuperscript{-1}.
They are comprised of non-equilibrium displacements of the atoms along the \textbf{y} and \textbf{z} crystallographic directions. The \begin{math}\sim\end{math}8 cm\textsuperscript{-1} optical phonon is a uniform in-plane shift of the atoms, estimated to be \begin{math}\sim\end{math}350 fm using a \begin{math}\sim\end{math}230 $\upmu$J/cm\textsuperscript{2} absorbed pump fluence, while the \begin{math}\sim\end{math}80 cm\textsuperscript{-1} mode corresponds to atomic displacements of few tens of femtometers which depend on the specific atom.
A more detailed description will be given further on in this work.

\begin{figure}[t]
 \includegraphics[width=\columnwidth]{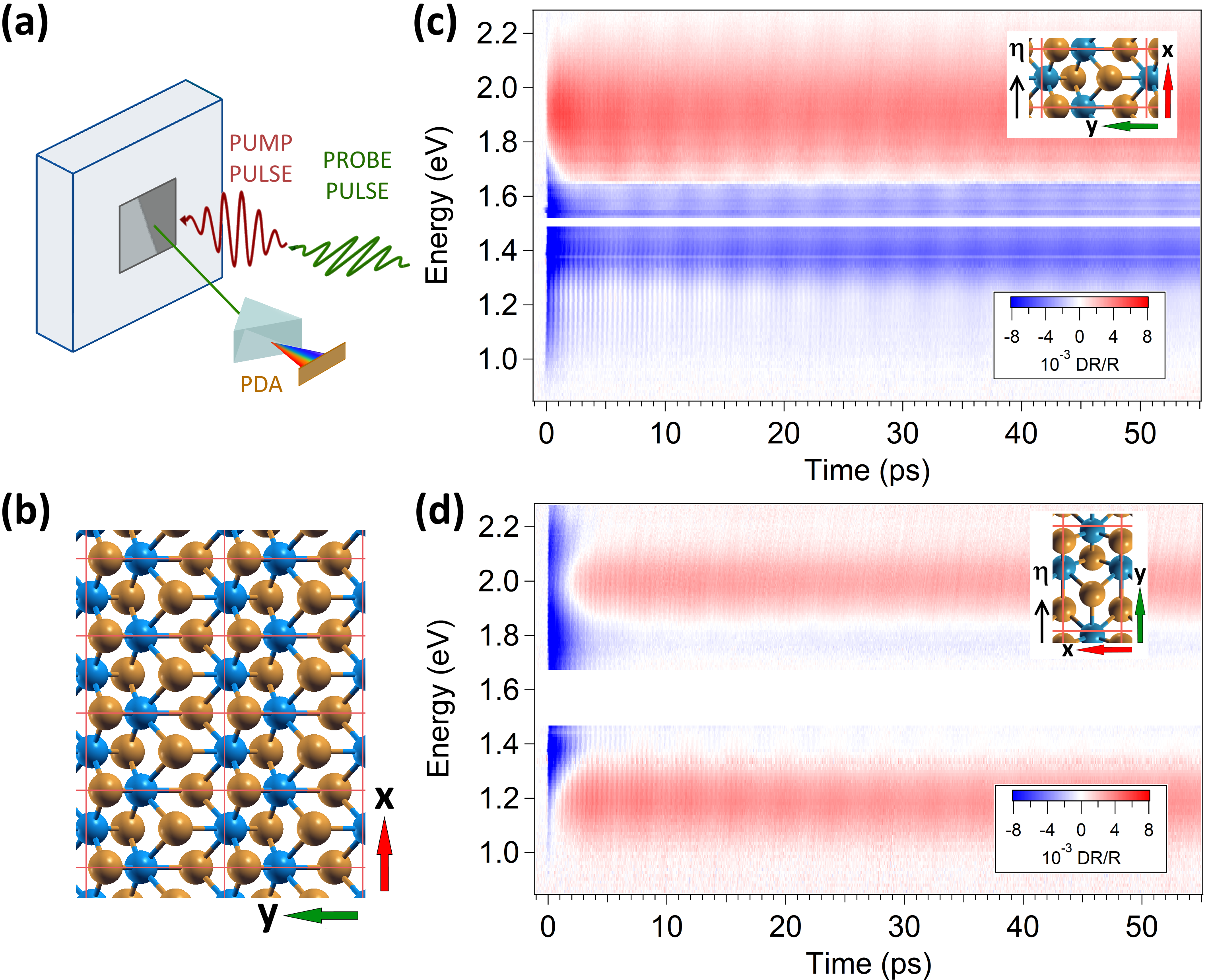}

\caption{(a) Sketch of the time-resolved broadband reflectivity experiment; the reflected probe beam is dispersed through a glass prism and detected by a photodiode array (PDA) detector. (b) Projection of a WTe\textsubscript{2} layer on the xy plane with blue tungsten and orange tellurium atoms showing multiple unit cells delimited by red lines; other projections are reported in \cite{SI}. (c),(d) Ultrafast broadband DR/R images for WTe\textsubscript{2} with the probe beam polarization along (c) \textbf{x} and (d) \textbf{y}, both taken at \begin{math}\sim\end{math}710 $\upmu$J/cm\textsuperscript{2} absorbed pump fluence and 1.55 eV pump photon energy at T=295 K.}

 \label{broadband295}
 \end{figure}

High-quality tungsten ditelluride samples were grown as reported in \cite{magnetoWTe2}. The presence of defects was previously studied \cite{magnetoWTe2}\cite{Das2016} and has negligible impact on our estimate of the average displacements.
In order to identify the in-plane crystallographic axes, LEED images were acquired under ultra-high vacuum conditions.
Time-resolved reflectivity experiments (sketch of the set-up in Fig. \ref{broadband295}(a)) were performed using a Ti:sapphire femtosecond (fs) laser system, delivering, at a repetition rate of 250 kHz, \begin{math}\sim\end{math}50 fs light pulses at a wavelength of 800 nm (1.55 eV).  The broadband (0.8-2.3 eV) supercontinuum probe beam was generated using a sapphire window.

DFT simulations were carried out using norm-conserving (NC) \cite{ncoriginal} scalar relativistic \cite{Takeda1978} pseudopotentials with the generalized gradient approximation (GGA) in the Perdew-Burke-Ernzerhof (PBE) parametrization for the exchange-correlation functional \cite{pbe} chosen from the PseudoDojo database \cite{onvcsg15}\cite{pseudodojo}. 
Structural optimizations and zone-center optical phonon calculations in the framework of density functional perturbation theory (DFPT) were performed using the \textsc {Quantum Espresso} (QE) \cite{QE-2009} suite of codes.
We calculated the diagonal macroscopic dielectric tensor components through the \textsc{Yambo} code \cite{YAMBO} at the independent-particle (IP) level, starting from the calculated wavefunctions and eigenvalues obtained through QE \cite{SI}.

The normalized time-resolved differential reflectivity (DR/R), measured in a near-normal incidence geometry, revealed a large anisotropy consistent with the two-fold in-plane symmetry of WTe\textsubscript{2} (Fig. \ref{broadband295}(b)) also reported for equilibrium electrodynamics \cite{anieq}.

In Figs. \ref{broadband295}(c),(d), the broadband probe beam polarization was set parallel to the \textbf x (\begin{math}\mathbf{\eta} || \mathbf{x}\end{math})  and \textbf{y} (\begin{math}\mathbf{\eta} || \mathbf{y}\end{math}) crystallographic axes respectively with pump beam polarization kept perpendicular to the probe one.
Measurements were acquired with a \begin{math}\sim\end{math}710 $\upmu$J/cm\textsuperscript{2} absorbed pump fluence and 1.55 eV pump photon energy at T=295 K.
A region of the spectrum was disturbed by the scattering of pump beam photons from the sample, so it was removed.

The temporal evolution of the reflectivity is directly related to the electronic and ionic degrees of freedom \cite{Dekorsy2000}\cite{Cheng}\cite{coherentphonons}. 
In general, the DR/R is a superposition of signals due to transient variations of electronic density of the states and population. We describe the temporal evolution after the perturbation (time-zero, t=0) as
\begin{multline}\label{drrgen}
\frac{DR}{R}(t, h\nu)=G(t) \ast \big[\sum_i A_ i(h\nu) e^{-t/\tau _ {Ri}}+\\
B(h\nu) e^{-t/\tau _ {heat}} + \sum_j C_ j(h\nu) cos(\omega_ j t+\phi_ j)e^{-t/\tau _{Pj}}\big]
\end{multline}

where \begin{math}G(t)\end{math} represents the pump and probe pulses cross-correlation and \begin{math}A_ i( h\nu)\end{math}, \begin{math}B(h\nu)\end{math} and \begin{math}C_j( h\nu)\end{math} denote the amplitudes of three different phenomena: i) electronic relaxation phenomena with time constants \begin{math}\tau _ {Ri}\end{math}, ii)  heating contribution with a characteristic time \begin{math}\tau_ {heat}\end{math} and iii) oscillations due to coherent phonons with angular frequency \begin{math}\omega_ j\end{math}, initial phase \begin{math}\phi_ j\end{math} and decay time \begin{math}\tau_{ Pj}\end{math}.

After the pump pulse excitation, electron-electron and electron-phonon scattering processes constitute the main incoherent relaxation phenomena in the first hundreds of femtoseconds and are responsible for the initial exponential decay of the DR/R signal. 
After a few picoseconds, the differential reflectivity DR/R reached a plateau, showcasing an offset with respect to the equilibrium value. It vanished in almost one nanosecond \cite{SI}. Due to its timescale, this effect is likely to be associated with local lattice heating, as described in \cite{latticeheating} for transition metals. For our purposes, we can consider the offset as a constant over the temporal range explored in Figs. \ref{broadband295}(c),(d).

The temporal profiles are very well fitted by using Eq. (\ref{drrgen}) \cite{SI}. Their sign and magnitude vary with the photon energy, while the main relaxation presents a time constant which is almost the same and equal to \begin{math}\sim\end{math}1 ps with only small differences, as previously found in \cite{Dai} at h\begin{math}\nu\end{math}=1.55 eV. 

 \begin{figure}[t]
 \includegraphics[width=\columnwidth]{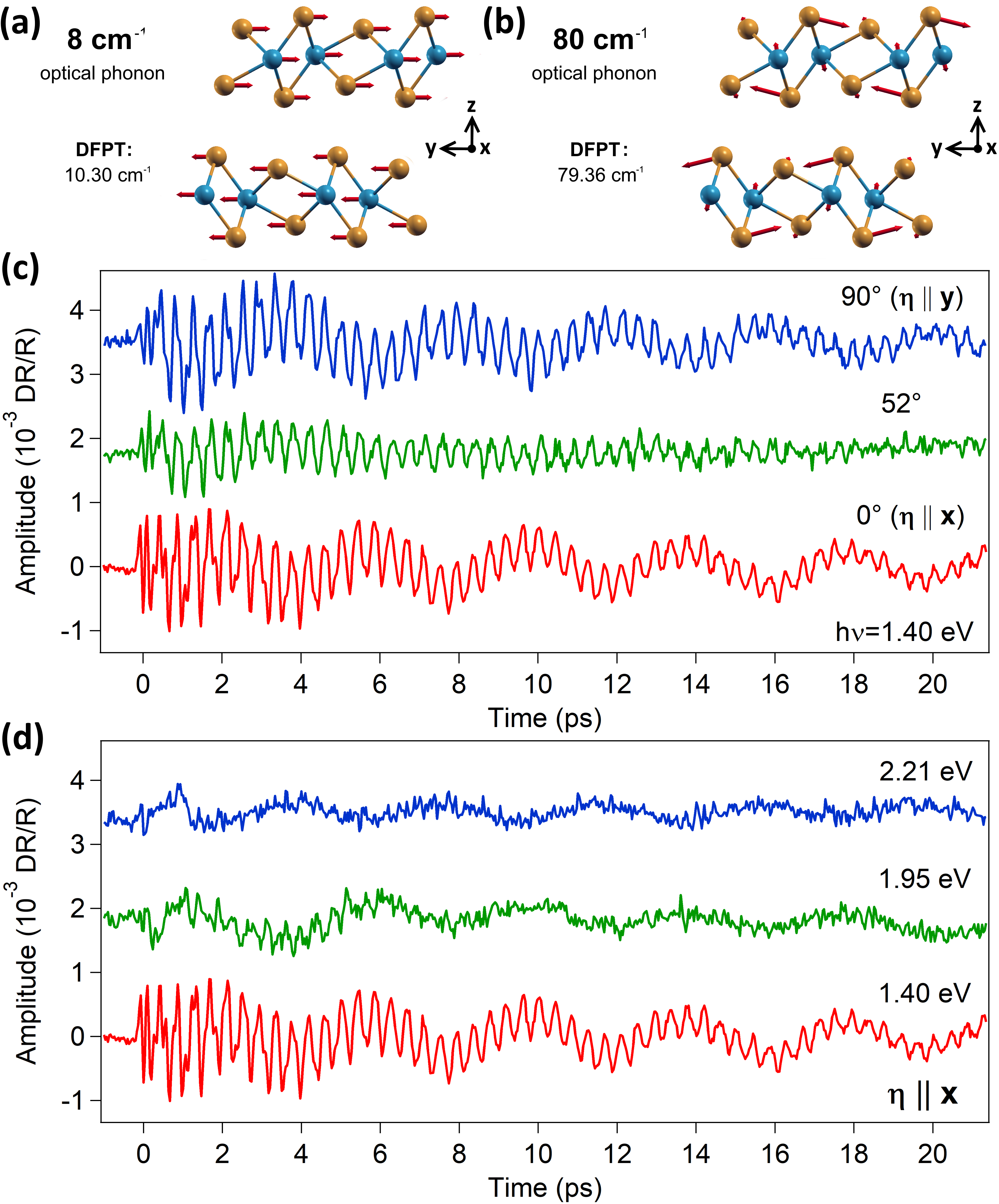}
\caption{DFPT eigendisplacements and frequencies for the (a) \protect 8 cm\textsuperscript{-1} and (b) \protect80 cm\textsuperscript{-1} optical modes; the tungsten atoms are blue, whereas the tellurium ones are orange. (c),(d) Oscillatory signal due to coherent phonons (the two upper curves are shifted from the central DR/R=0) (c) at 1.40 eV photon energy for different probe beam polarization angles, showing a \begin{math}\pi\end{math} phase shift for 8 cm\textsuperscript {-1} mode between the two perpendicular configurations while the 80 cm\textsuperscript {-1} mode appears to be unaffected and (d) for \begin{math}\mathbf{\eta} || \mathbf{x}\end{math} as a function of the photon energy, evidencing a \begin{math}\pi\end{math} phase shift for 8 cm\textsuperscript {-1} mode when comparing the infrared (1.40 eV) to the visible (2.21 eV) spectral region.}
 \label{fig:phonons295}
 \end{figure}
 
Subtracting an exponential fit function from the experimental signals allows the extraction of the coherent component of the DR/R signal. 
The resulting signal displays marked periodic modulations in the time domain, arising from the excitation of coherent phonon modes. The most prominent oscillating features, as detected by Fast Fourier Transform (FFT) and in \cite{Dai}, have frequency of 7.9\textpm0.4 cm\textsuperscript {-1} and 79.7\textpm 0.4 cm\textsuperscript {-1}. For notation purposes, we refer to these modes with the labels 8 cm\textsuperscript{-1} and 80 cm\textsuperscript{-1}.
The associated time-decay constants, derived by fitting the modulations with two exponentially-damped cosine waves, are \begin{math}\tau_{P1}\end{math}=77\begin{math}\pm\end{math}4 ps for the 8 cm\textsuperscript {-1} mode and \begin{math}\tau_{P2}\end{math}=12.1\begin{math}\pm\end{math}0.2 ps for the 80 cm\textsuperscript {-1} mode. DFPT results indicate that these frequencies are linked to A\textsubscript{1} zone-center optical phonon modes,  represented in Fig. \ref{fig:phonons295}(a),(b). The 8 cm\textsuperscript{-1} mode can be pictured as adjacent layers moving in antiphase, while the 80 cm\textsuperscript{-1} involves more complex in-plane and out-of-plane movements. The smaller value of the experimental frequency for the 8 cm\textsuperscript{-1} mode with respect to the DFPT result (10.30 cm\textsuperscript {-1}) can be attributed to a redshift as temperature or fluence are increased \cite{slowph}.
The marked difference between the time constants of the two modes could be linked to the different type of perturbation of the interatomic bonds induced by the associated displacements. The 8 cm\textsuperscript{-1} mode (Fig. \ref{fig:phonons295}(a)) only alters the interplanar, mainly van der Waals, interactions. Differently, the 80 cm\textsuperscript{-1} mode (Fig. \ref{fig:phonons295}(b)) induces a modification of the in-plane covalent bonds. A more detailed DFPT analysis could provide a rigorous basis for this intuitive argument, which is beyond the scope of the present work.

Weaker additional A\textsubscript{1} contributions at 116.5\textpm 0.4 cm\textsuperscript {-1}, 132.2\textpm 0.4 cm\textsuperscript {-1} and 210.2\textpm 0.4 cm\textsuperscript {-1} frequency appear as beats in the DR/R signal in the first hundreds of femtoseconds.

\begin{figure}[t]
 \includegraphics[width=\columnwidth]{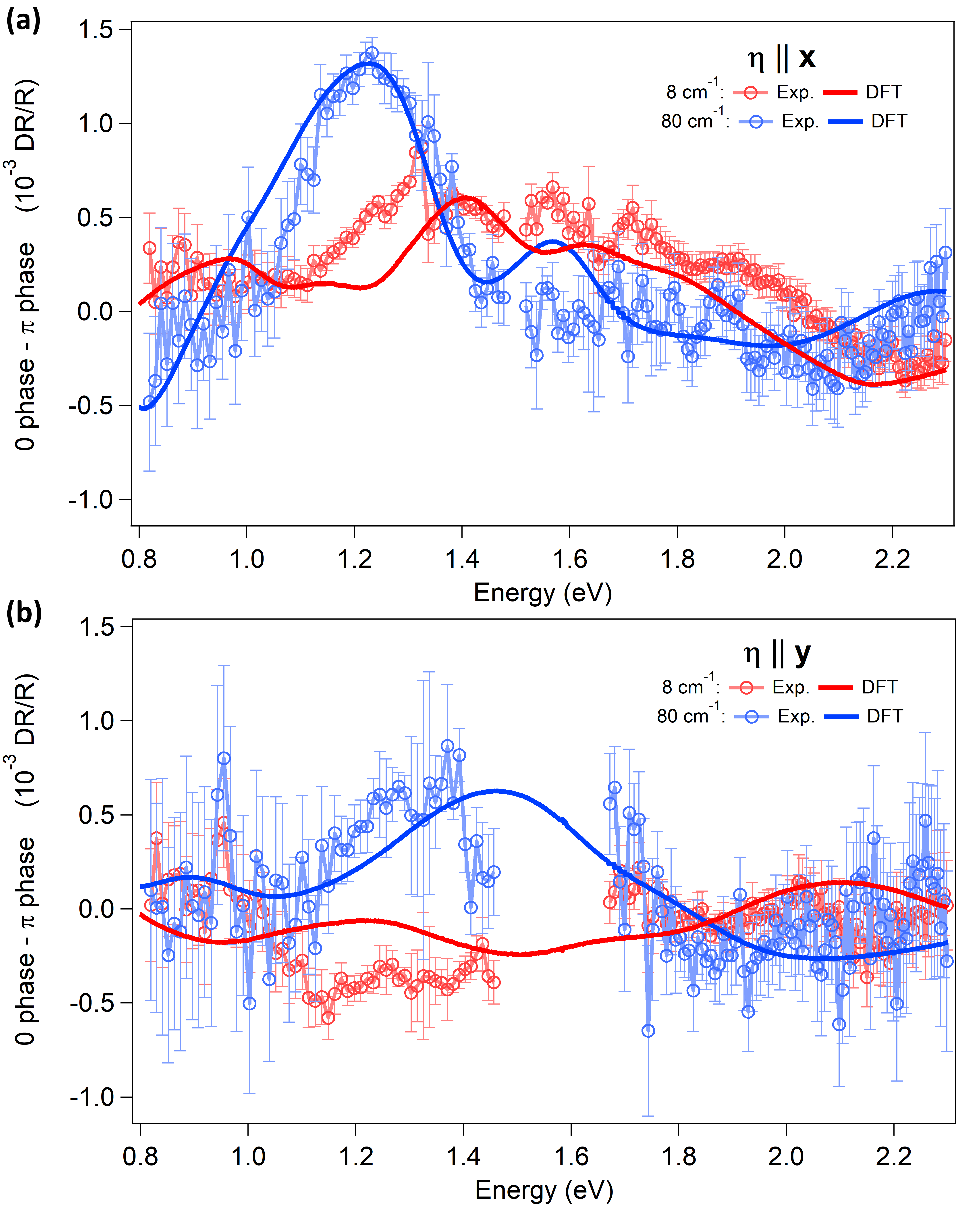}

\caption{Comparisons between the experimental and DFT calculated DR/R difference between the 0 and \protect\begin{math}\pi\protect\end{math} phases, showcasing the effects of the 8 cm\textsuperscript {-1} and 80 cm\textsuperscript {-1} optical phonons for (a) \protect\begin{math}\mathbf{\eta} || \mathbf{x}\protect\end{math} and (b) \protect\begin{math}\mathbf{\eta} || \mathbf{y}\protect\end{math} using a \begin{math}\sim\end{math}710 $\upmu$J/cm\textsuperscript{2} absorbed pump fluence and T=295 K.}
\label{fig:comparison2}
 \end{figure}

When switching the probed direction from \begin{math}\mathbf{\eta} || \mathbf{x}\end{math} to \begin{math}\mathbf{\eta} || \mathbf{y}\end{math}, a \begin{math}\pi\end{math} phase change for the 8 cm\textsuperscript{-1} mode (Fig. \ref{fig:phonons295}(c)) was clearly registered for most of the probe photon energies.
Analogous \begin{math}\pi\end{math} phase changes were found by comparing the temporal profiles taken at different probe photon energies using the same polarization for the same mode (Fig. \ref{fig:phonons295}(d)). These phase differences can be explained in terms of the peculiar anisotropy of the dielectric function and were reproduced through our numerical simulations (see discussion for Fig. \ref{fig:comparison2}).

In order to model the effects of the coherent phonons on the reflectivity, we first calculated the wavefunctions and eigenvalues for the equilibrium configuration using QE and the macroscopic dielectric tensor diagonal components and the reflectivity curves through \textsc{Yambo}. Then, we considered four out-of-equilibrium configurations labeled as 0 and \begin{math}\pi\end{math} phases, corresponding to displacements in opposite directions, for each of the coherent phonon modes, using the eigendisplacements obtained through DFPT with respect to the equilibrium positions. 
The optical phonons led to modifications of the electronic band structure that, although small, were beyond the numerical accuracy and were confirmed by linearly rescaling the effects of larger perturbations \cite{SI}.
For each configuration we calculated the associated reflectivity and DR/R with respect to the equilibrium configuration. These curves describe the effect of the optical phonons on a quasi-equilibrium adiabatic system. Indeed, a few picoseconds after time-zero, the system has relaxed through electron-phonon scattering processes and the incoherent part of the DR/R signal reaches a plateau having a decay time much larger than the coherent phonon period and damping timescale.

\begin{figure}[t]
 \includegraphics[width=\columnwidth]{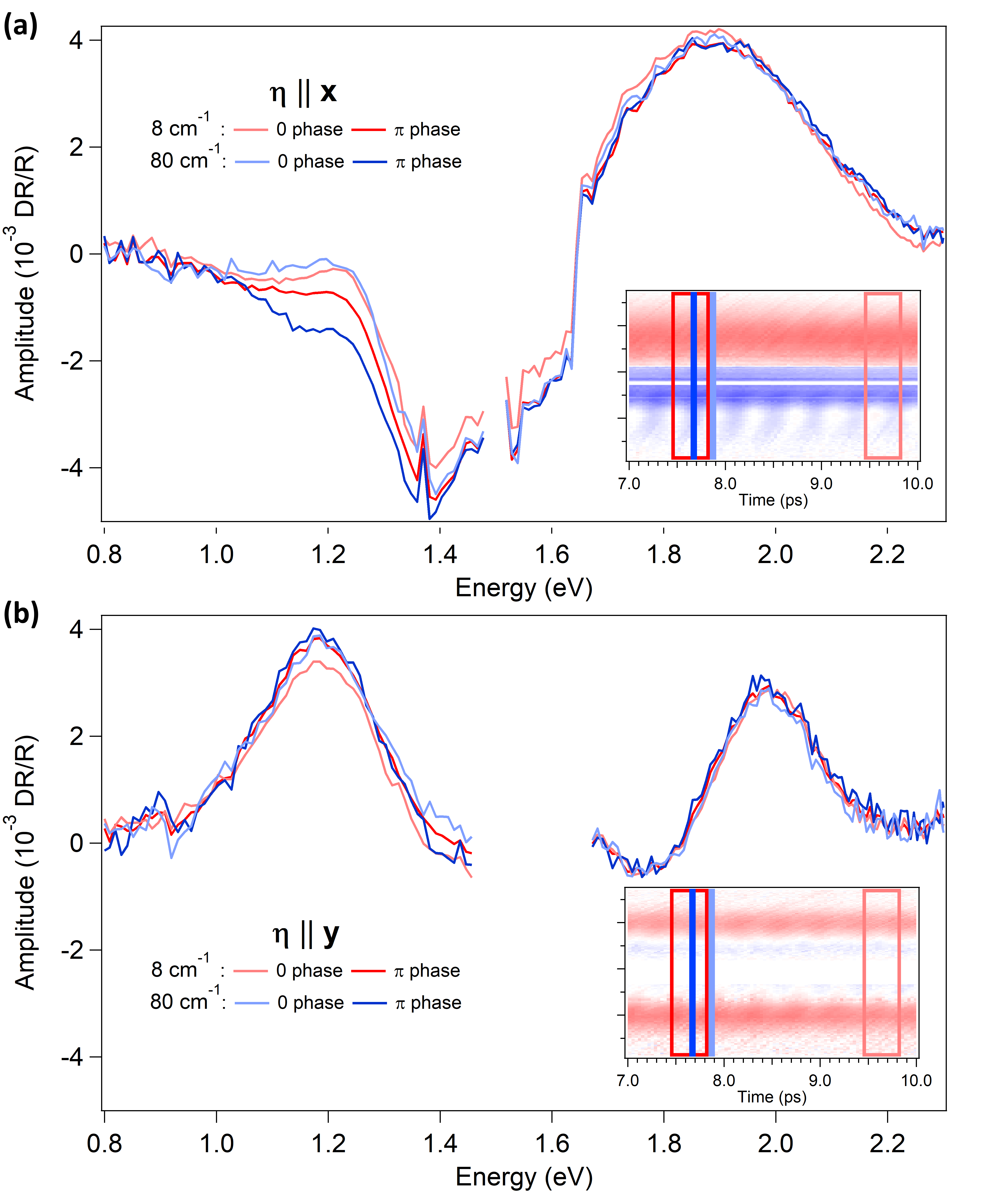}

\caption{Experimental spectral profiles extracted from the DR/R (t,h\textnu ) maps reported in Figs. \ref{broadband295}(c),(d). The linearly polarized probe is oriented for (a) \protect\begin{math}\mathbf{\eta} || \mathbf{x}\protect\end{math} and (b) \protect\begin{math}\mathbf{\eta} || \mathbf{y}\protect\end{math}. The 0 and \protect\begin{math}\pi\protect\end{math} phases are referred to the maximum and minimum of the oscillation in the DR/R signal for \protect\begin{math}\mathbf{\eta} || \mathbf{x}\protect\end{math} in the infrared spectral region; the insets show the time delays at which the curves were extracted. The vertical axis and color scale of the insets are the same as for Figs. \ref{broadband295}(c),(d).}
\label{fig:comparison1}
 \end{figure}

Experimentally, the effects of each single optical phonon mode on the DR/R were isolated with the following procedure. For the 8 cm\textsuperscript{-1} phonon, we took the mean value over one period of the 80 cm\textsuperscript{-1} modulation around two consecutive 8 cm\textsuperscript{-1} extrema (Fig. \ref{fig:comparison1}). For the 80 cm\textsuperscript{-1} oscillation, we directly used the spectral profiles at the extrema, considering that the period of the slower phonon is about ten times larger. The spectral profiles were collected after the main relaxation peak, as precisely indicated in the insets. The difference between the spectral profiles at the maximum (0 phase) and minimum (\begin{math}\pi\end{math} phase) of the coherent phonons, named with respect to the \begin{math}\mathbf{\eta} || \mathbf{x}\end{math} DR/R, gives an unique signature of the phonon mode.

Analogously, in our calculations, we considered the difference between the calculated DR/R curves at 0 and \begin{math}\pi\end{math}, multiplied by an exponential factor derived from the experimental data, to account for the damping of the optical modulation.
The DFT predictions closely match the experimental results for both the 8 cm\textsuperscript{-1} and 80 cm\textsuperscript{-1} modes by rescaling the atomic displacements, whose relative amplitude is given by DFPT, with a global multiplicative factor, common to both polarizations (Fig. \ref{fig:comparison2}).
The agreement is particularly good for \begin{math}\mathbf{\eta} || \mathbf{x}\end{math} where the phonon effects are most prominent. Peculiar features, such as their trend and change of sign, are maintained throughout the relaxation process as previously illustrated in Fig. \ref{fig:phonons295}(c),(d).

From this comparison we can quantify the magnitude of the ionic displacements right after the perturbation, when the atoms of different unit cells move uniformly in the probed region.
The estimated displacements have to be regarded as average displacements over the probed volume and photon energies.  The penetration depth, derived from the optical data in \cite{anieq}, is almost constant and similar for the two perpendicular polarizations in the probed spectral region, with minor deviations towards the infrared for \begin{math}\mathbf{y}\end{math} direction.  

In order to evaluate very small atomic displacements, we repeated the experiments at lower fluence (\begin{math}\sim\end{math}230 $\upmu$J/cm\textsuperscript{2}). We checked that the DR/R spectral shape does not change at lower temperature or lower fluence, where we showed that the phonon effects can be considered linear with the deposited energy \cite{SI}.
This is not the case for \begin{math}\sim\end{math}710 $\upmu$J/cm\textsuperscript{2}, since the phonon effects measured at that fluence are only twice those at \begin{math}\sim\end{math}230 $\upmu$J/cm\textsuperscript{2}.
At \begin{math}\sim\end{math}230 $\upmu$J/cm\textsuperscript{2}, for the 8 cm\textsuperscript{-1} mode, the initial positions shifts are found to be \begin{math}\sim\end{math}350 fm along the \begin{math}\mathbf{y}\end{math} direction. For the 80 cm\textsuperscript{-1} mode, the displacements are quite distinct for the twelve atoms in the basis cell. The four tungsten atoms remain almost fixed (displacements smaller than 10 fm) along the \begin{math}\mathbf{y}\end{math} direction whereas they shift by \begin{math}\sim\end{math}40 fm along the interlayer \textbf{z} direction. Four tellurium atoms are displaced by \begin{math}\sim\end{math}90 fm along the \begin{math}\mathbf{y}\end{math} direction and by \begin{math}\sim\end{math}25 fm along \begin{math}\mathbf{z}\end{math}. The remaining four tellurium atoms are displaced by \begin{math}\sim\end{math}15 fm along the \begin{math}\mathbf{y}\end{math} direction and by \begin{math}\sim\end{math}30 fm along \begin{math}\mathbf{z}\end{math}. None of these modes involves movements along the \begin{math}\mathbf{x}\end{math} crystallographic direction.

We highlight that with the present approach the atomic displacements can be evaluated with a precision of a few femtometers without free tuning parameters, except a scaling factor determined by an overall comparison between experimental data and numerical simulations of the broadband DR/R signal. Concerning WTe\textsubscript{2}, a good agreement between the measured and calculated phonon effects on the optical properties for both in-plane crystal axes has been obtained. This finding confirms that the macroscopic lattice distortions excited in WTe\textsubscript{2} at 8 cm\textsuperscript{-1} and 80 cm\textsuperscript{-1} can be entirely mimicked by the coherent population of selected zone-center A\textsubscript{1} lattice modes. We stress that the possibility to measure the optical properties on a wide spectral range is fundamental to extract reliable quantitative results about the magnitude of eigendisplacements of multiple and intertwined phonon modes. Finally, we mention that the method described here is not system-specific and in principle can be extended to any crystalline material, provided that its high-energy optical properties are affected by the coherent motion of atoms. In perspective, our findings can pave the way to the design of tailored devices in which the coherent lattice motion (at selected frequencies and up to large displacements) is exploited to finely tune the functional properties of semiconducting and metallic systems.

We acknowledge fruitful discussion and suggestions concerning the use of the \textsc{Yambo} code by Davide Sangalli and Andrea Marini. We thank Paolo Giannozzi for useful discussions.
D.S. acknowledges support from the European Social Fund Operational Programme 2014/2020 of Region Friuli Venezia Giulia.
High-performance computing resources were obtained from CINECA through the ISCRA initiative and the agreement with the University of Trieste and from Elettra Sincrotrone Trieste.
R.J.C. acknowledges that funding for the growth of the crystals used in this work was provided by the US NSF MRSEC program, grant DMR-1420541.
F.P. and M.P. acknowledge the University of Trieste for supporting this research with the projects FRA 2016, code J96J16000980005 and FRA 2018, code J91G18000730002, respectively.

\bibliography{MTa}

\end{document}


\captionsetup[figure]{labelfont={default},labelformat={default},labelsep=period,justification={raggedright}}
\renewcommand{\thefigure}{S\arabic{figure}}

\title{Supplemental material for \\
 Ultrafast broadband optical spectroscopy for quantifying subpicometric coherent atomic displacements in WTe\textsubscript{2}}

\author{Davide Soranzio}

\affiliation{Dipartimento di Fisica, Universit\`a degli Studi di Trieste, 34127 Trieste, Italy}
\author{Maria Peressi}
\affiliation{Dipartimento di Fisica, Universit\`a degli Studi di Trieste, 34127 Trieste, Italy}
\author{Robert J. Cava}
\affiliation{Department of Chemistry, Princeton University, Princeton, New Jersey 08544, USA}
\author{Fulvio Parmigiani}
\affiliation{Dipartimento di Fisica, Universit\`a degli Studi di Trieste, 34127 Trieste, Italy}
\affiliation{International Faculty, University of Cologne, Albertus-Magnus-Platz, 50923 Cologne, Germany}
\affiliation{Elettra-Sincrotrone Trieste S.C.p.A., 34149 Basovizza, Italy}
\author{Federico Cilento}
\email[e-mail:]{federico.cilento@elettra.eu}
\affiliation{Elettra-Sincrotrone Trieste S.C.p.A., 34149 Basovizza, Italy}

\date{\today}

\maketitle

\section{Part 1 - Crystal structure of tungsten ditelluride}
\begin{figure*}[h!]
 \includegraphics[width=\textwidth]{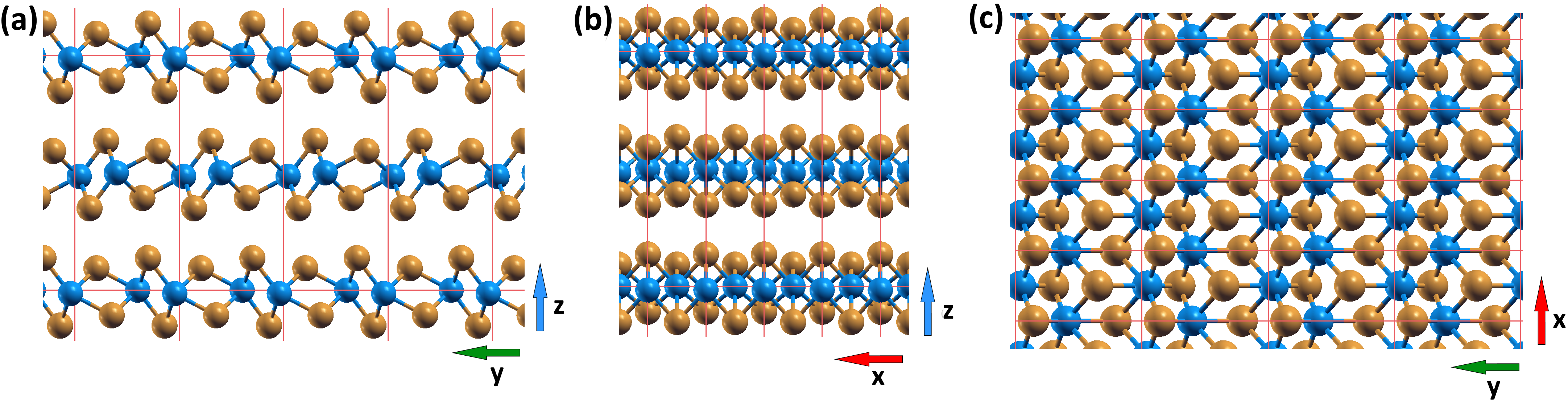}

\caption{Projections of WTe\textsubscript{2} crystal structure perpendicular to the (a) \textbf{x}, (b) \textbf{y} and (c) \textbf{z} (only a single layer is shown) axes; \textbf{x} and \textbf{y} are the perpendicular in-plane directions, while the \textbf{z} axis is perpendicular to the layers; tungsten atoms in blue, tellurium atoms in orange; the unit cells are delimited by red lines.}

 \label{fig:projection}
 \end{figure*}

In Fig. \ref{fig:projection}, ball-and-stick projections of the crystal structure of the layered orthorhombic phase of tungsten ditelluride (WTe\textsubscript{2}) are shown. The crystallographic data were taken from \cite{cryst}.  
All the crystallographic representations were generated using the XCrySDen \cite{xcrysden} software.

\section{Part 2 - Experimental results}
All the experiments were executed at the T-ReX laboratory at Elettra Sincrotrone Trieste.
The time-resolved reflectivity experiments were performed using a Ti:sapphire femtosecond (fs) laser system, delivering, at a repetition rate of 250 kHz, \begin{math}\sim\end{math}50 fs light pulses at a wavelength of 800 nm (1.55 eV). 

The output beam of the laser system was split into pump and probe beams which reunite at the sample. During their paths, the pump beam was delayed through a linear stage to control the pump-probe delay, while the probe beam was focused in a sapphire window to produce a broad (0.8-2.3 eV) supercontinuum spectrum.
On the sample, the pump spot diameter was estimated to be \begin{math}\sim\end{math}160 $\upmu$m, while the average spot diameter for the various spectral components of the probe beam was \begin{math}\sim\end{math}110 $\upmu$m.
After reflection from the sample in quasi-normal conditions, the probe beam's spectral components are dispersed through a glass prism and measured with an InGaAs photodiode array (PDA).

Modifying the pump polarization direction produced no significant variation of the main spectral features. For this reason, all  measurements were performed by keeping the pump beam polarization perpendicular to the probe one, in order to minimize the residual pump scattering by using a polarizer inserted before the detectors.

Error intervals are expressed as \textpm \textsigma, one standard deviation. 
\subsection{2.1 - Measurement at 1.03 eV pump photon energy}

\begin{figure*}[t]
 \includegraphics[width=\textwidth]{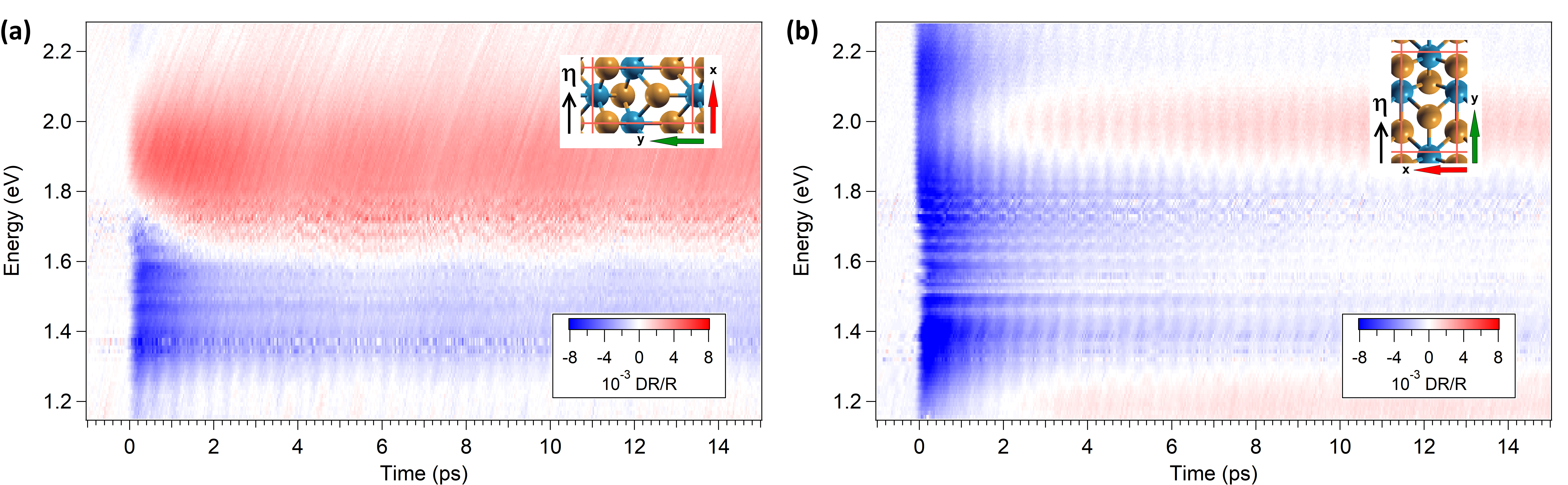}

\caption{(a),(b) Data obtained with a 1.03 eV pump photon energy, taken at \protect\begin{math}\sim\protect\end{math}490 $\upmu$J/cm\textsuperscript{2} absorbed pump fluence and T=295 K respectively for \protect\begin{math}\mathbf{\eta} || \mathbf{x}\protect\end{math} and \protect\begin{math} \mathbf{\eta} || \mathbf{y}\protect\end{math}.}

 \label{fig:295mix}
 \end{figure*}

In order to check that the reflectivity variation DR/R is independent from the pump photon energy, we repeated part of the measurements by pumping at 1.03 eV photon energy (Fig. \ref{fig:295mix}), obtained thanks to an optical parametric amplifier (OPA). We verified that the DR/R is qualitatively similar to the results obtained with the 1.55 eV (800 nm) excitation. Moreover, since in this experimental condition the pump lies at the edge of the measured spectral range, the spectral region around 1.55 eV is not covered by scattered pump photons. This allowed us to confirm that no structured features are present in this range, that was hidden in the datasets of the main paper acquired with 1.55 eV pump photon energy (Figs. 1(c),(d), main paper).

 \subsection{2.2 - Time-domain analysis}

 \begin{figure*}[t]
 \includegraphics[width=\textwidth]{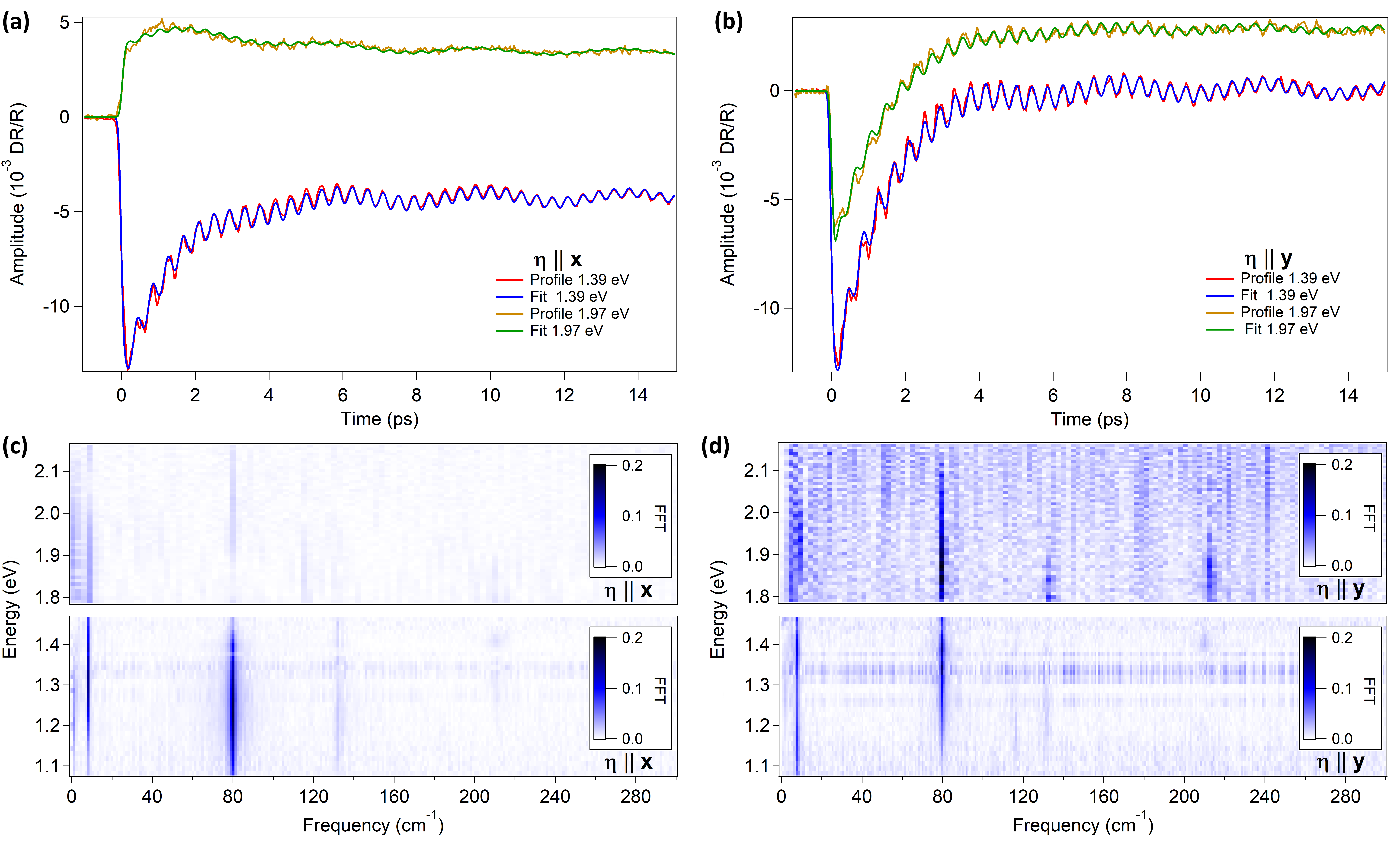}

\caption{Profiles at 1.39 eV and 1.97 eV photon energy for probe beam polarization (a) \begin{math}\mathbf{\eta} || \mathbf{x}\end{math} and (b)  \begin{math}\mathbf{\eta} || \mathbf{y}\end{math}, at \begin{math}\sim\end{math}710 \textmu J/cm\textsuperscript{2} absorbed pump fluence and T=295 K. Fast Fourier Transform images performed across a few picoseconds after time zero for (c) \begin{math}\mathbf{\eta} || \mathbf{x}\end{math} and (d) \begin{math}\mathbf{\eta} || \mathbf{y}\end{math} after subtracting from the data a purely exponential fit based on Eq. (1) in the main paper for various probe photon energies.} 

 \label{fig:295profilesFFT}
 \end{figure*}
  \begin{figure*}[h!]
 \includegraphics[width=\textwidth]{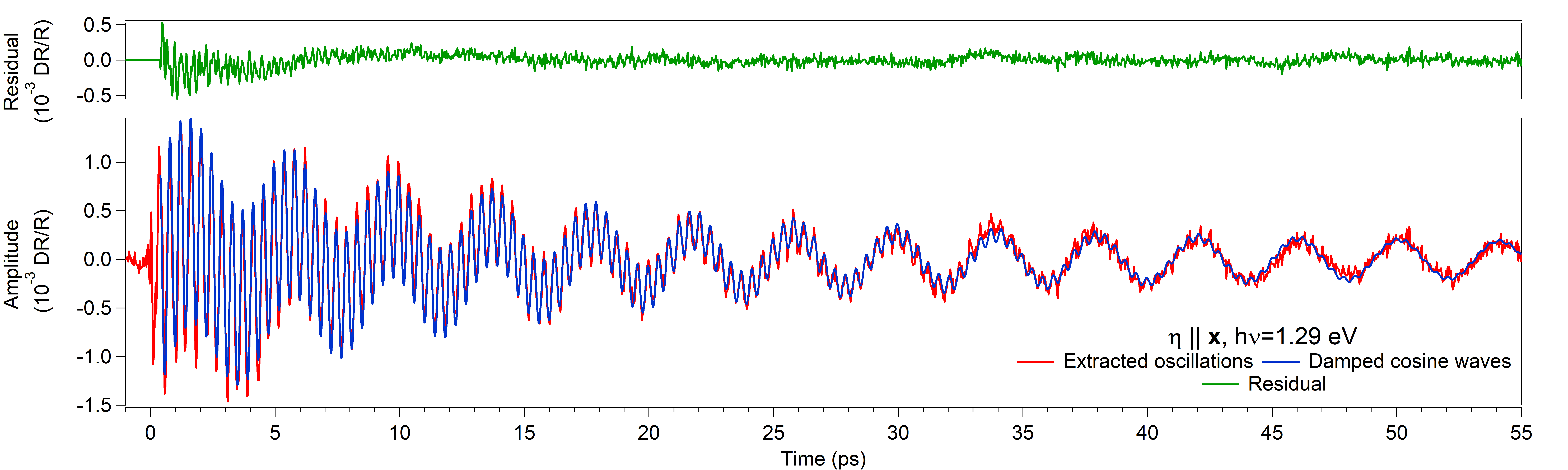}

\caption{Oscillatory component of the DR/R signal, obtained by subtracting an exponential fit (see Eq. (1) in the main paper) from the temporal profile at 1.29 eV acquired at \begin{math}\sim\end{math}710 \textmu J/cm\textsuperscript{2} absorbed pump fluence and T=295 K for \begin{math} \mathbf{\eta} || \mathbf{x}\end{math}.}

 \label{fig:295oscifit}
 \end{figure*}
\begin{figure*}[t]
 \includegraphics[width=\textwidth]{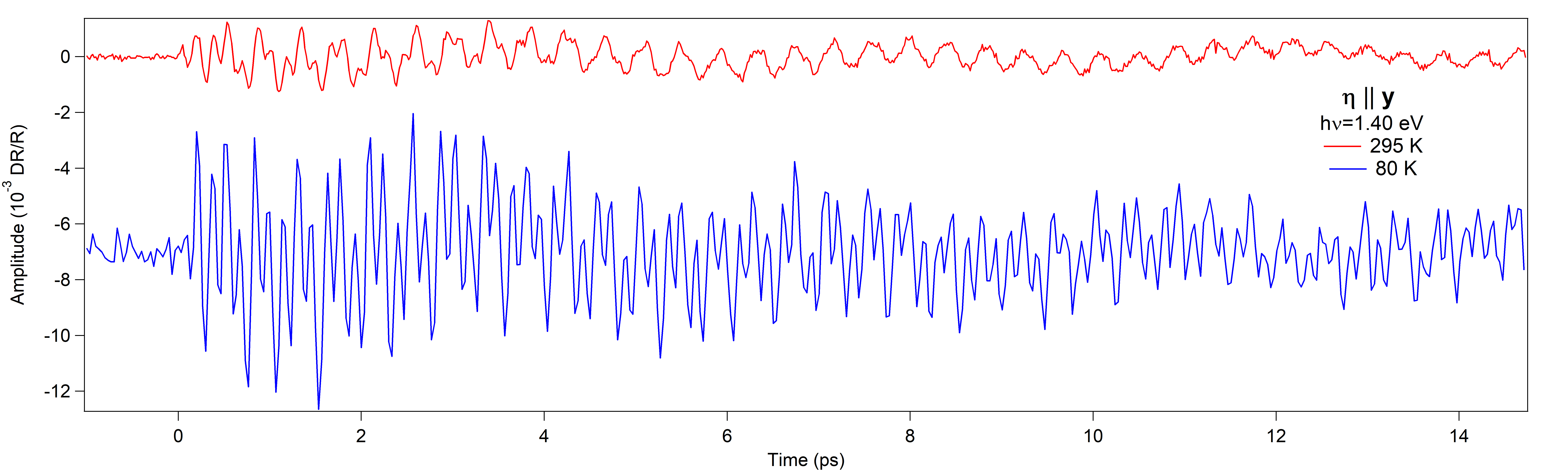}

\caption{Phonon modulation of DR/R, obtained by subtracting an exponential fit (see Eq. (1) in the main paper) from the temporal profiles at 1.40 eV acquired at \begin{math}\sim\end{math}710 \textmu J/cm\textsuperscript{2} absorbed pump fluence, T=80 K and T=295 K and for \begin{math} \mathbf{\eta} || \mathbf{y}\end{math}; the beats are due to phonon modes with frequency higher than 80 cm\textsuperscript{-1}.}

 \label{fig:16supp}
 \end{figure*}

 In Fig. \ref{fig:295profilesFFT} we report selected profiles extracted at two photon energies from the datasets reported in Figs. 1(c),(d) in the main paper, alongside with the Fast Fourier Transform (FFT) of the oscillatory signal in most of the investigated range. The most prominent features are at \begin{math}\sim\end{math}8 and 80 cm\textsuperscript{-1}. These modes give the largest contribution to the coherent component of the DR/R signal as shown in Fig. \ref{fig:295oscifit}, where two damped cosine waves fit the oscillatory signal. Minor contributions at higher frequencies (\begin{math}\sim\end{math}117, 132 and 210 cm\textsuperscript {-1}) are detected and appear as beats in the oscillatory signal. Their contribution becomes more relevant as the temperature is reduced (Fig. \ref{fig:16supp}).

\begin{figure*}[t]
 \includegraphics[width=\textwidth]{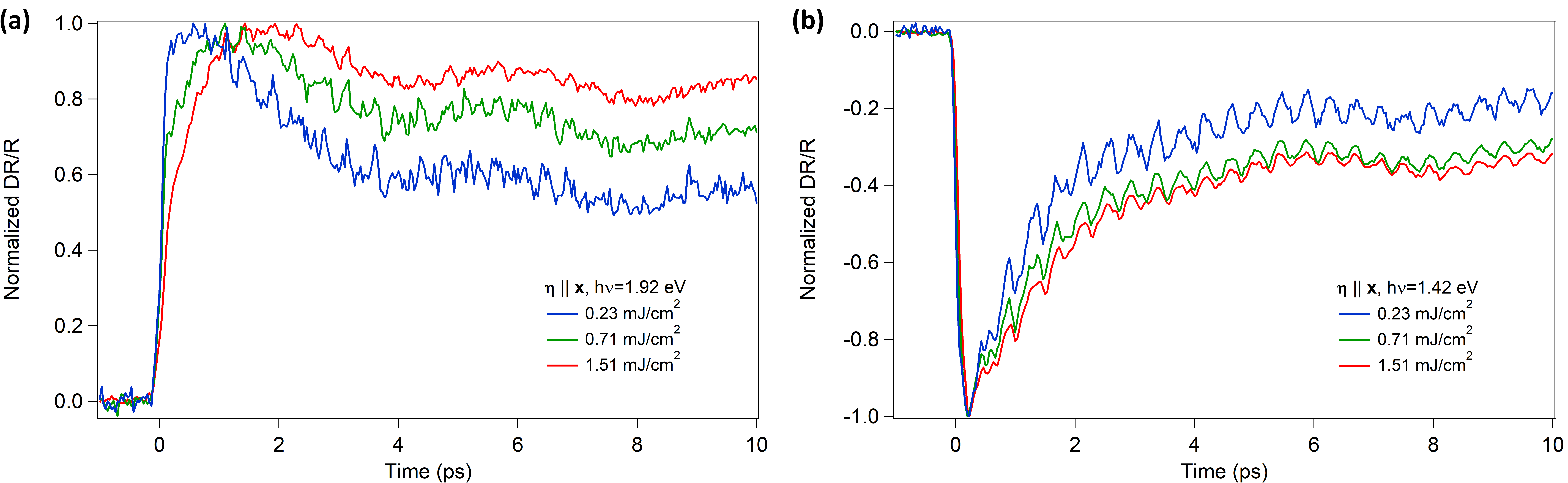}

\caption{DR/R profiles obtained  from 1.55 eV pump photon energy-broadband supercontinuum probe beam experiments at different fluences and T=295 K for probe polarization \protect\begin{math}\mathbf{\eta} || \mathbf{x}\protect\end{math}; the profiles in (a) at 1.92 eV show a fluence-dependent position of the main out-of-equilibrium peak, while the profiles in (b) at 1.42 eV do not present such behavior.}
 \label{fig:peakshift}
 \end{figure*}

In the visible spectral region (1.70-2.10 eV), the DR/R cannot be fitted by a single exponential decay (Eq. (1) in the main paper) for probe polarization \begin{math}\mathbf{\eta} || \mathbf{x}\end{math}.
In presence of contributions with opposite sign and comparable time constants, the DR/R reaches its maximum value at longer delay, analogously to the case in Fig. \ref{fig:295profilesFFT}(a). This delay can be as large as 1 ps, alongside with a slower decay of the signal.  Investigating the dependence on fluence, we learned that (Fig. \ref{fig:peakshift}) the delay increases with the deposited pump energy.
At variance, in the infrared the peak position is independent on the fluence. For this reason, for example, it is not possible to effectively describe the behavior of this material with just a single Drude-Lorentz term \cite{wooten} in the dielectric tensor for the measured spectral range. For \begin{math}\mathbf{\eta} || \mathbf{y}\end{math}, we do not observe such phenomenon. This may be connected to different matrix elements for the involved transitions in the two distinct linear probe polarizations.

 \subsection{2.3 - Low fluence and low temperature measurements}
\begin{figure*}[h!]
 \includegraphics[width=\textwidth]{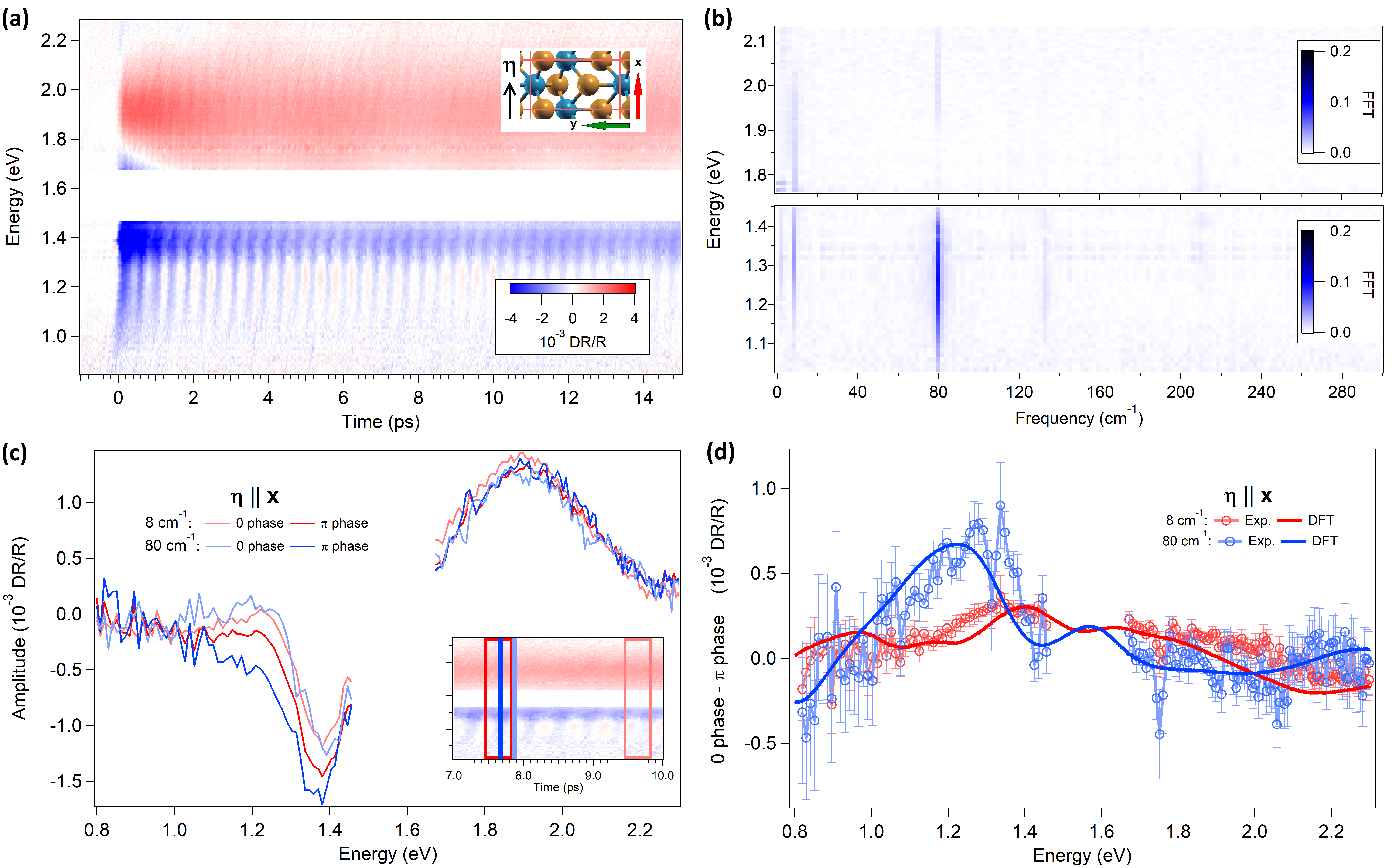}
\caption{(a) Ultrafast broadband DR/R image for WTe\textsubscript{2} for probe polarization \protect\begin{math}\mathbf{\eta} || \mathbf{x}\protect\end{math} taken at \protect\begin{math}\sim\protect\end{math}230 $\upmu$J/cm\textsuperscript{2} absorbed pump fluence and T=295 K. (b) Fast Fourier Transform images performed across a few picoseconds after time zero after subtracting from the data a purely exponential fit based on Eq. (1) in the main paper. (c) Spectral profiles taken at selected time delays for maxima and minima of the 8 cm\textsuperscript{-1} and 80 cm\textsuperscript{-1} optical phonon effects; the inset, with the same vertical axis and color scale as (a), shows the exact temporal positions. (d) Comparisons between the experimental and DFT calculated DR/R difference between the 0 and \protect\begin{math}\pi\protect\end{math} phases displayed in (c), showcasing the effects of the 8 cm\textsuperscript {-1} and 80 cm\textsuperscript {-1} optical phonons for \protect\begin{math}\mathbf{\eta} || \mathbf{x}\protect\end{math}.}
 \label{fig:lowfluence}
 \end{figure*}

The DR/R broadband image measured at \begin{math}\sim\end{math}230 $\upmu$J/cm\textsuperscript{2} absorbed pump fluence and T=295 K (Fig. \ref{fig:lowfluence}) presents the same characteristics of that measured at higher fluence (see Figs. 1(c), 3(a) and 4(a) in the main paper). Such features are also well captured by ab-initio results obtained with properly rescaled eigendisplacements.

\begin{figure*}[h!]
 \includegraphics[width=\textwidth]{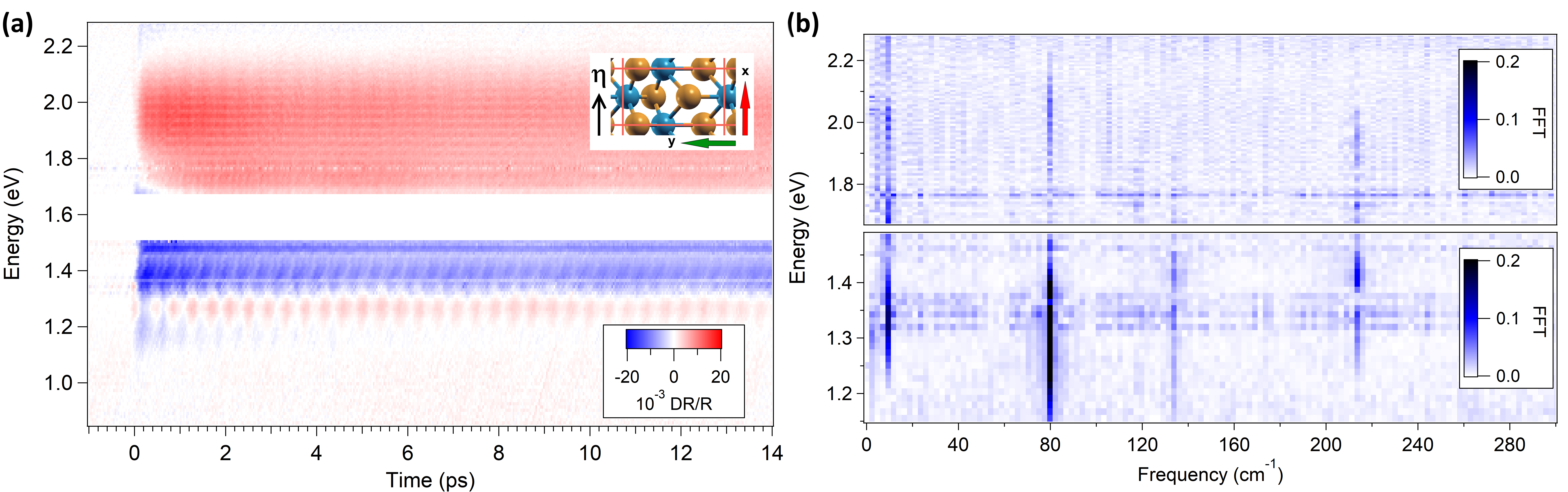}
 \caption{(a) Ultrafast broadband DR/R image for WTe\textsubscript{2} for probe polarization \protect\begin{math}\mathbf{\eta} || \mathbf{x}\protect\end{math} taken at \protect\begin{math}\sim\protect\end{math}710 $\upmu$J/cm\textsuperscript{2} absorbed pump fluence and T=80 K. (b) Fast Fourier Transform images performed across a few picoseconds after time zero after subtracting from the data a purely exponential fit based on Eq. (1) in the main paper for various probe photon energies.}
 \label{fig:80Kahigh}
 \end{figure*}

\begin{figure*}[h!]
 \includegraphics[width=\textwidth]{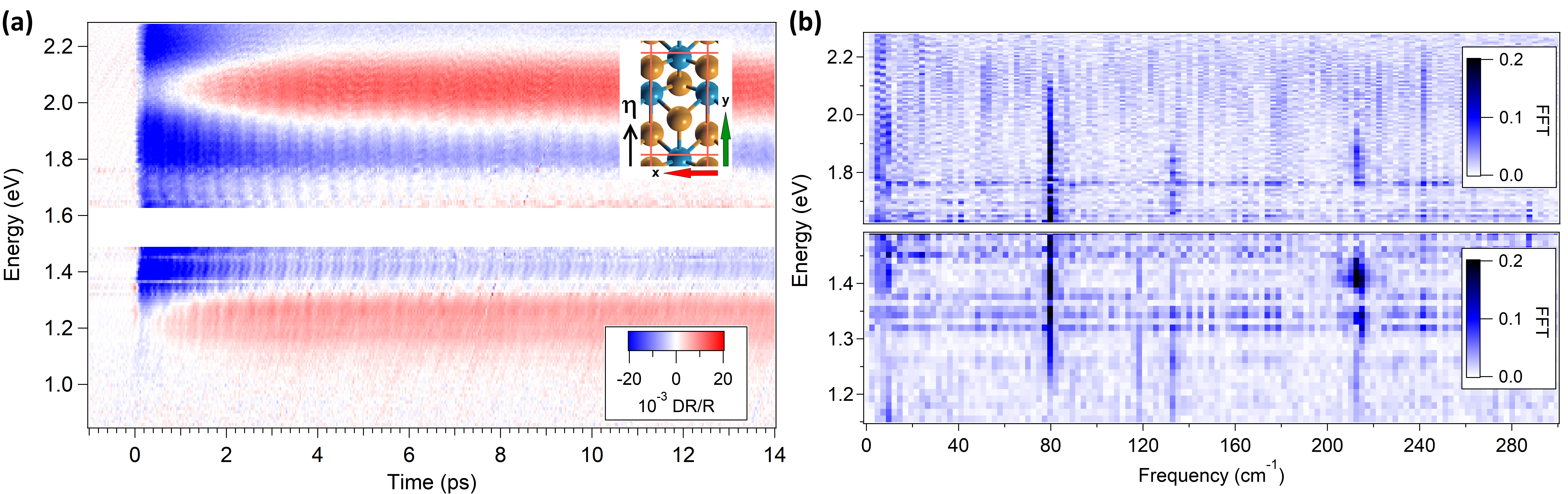}
 \caption{(a) Ultrafast broadband DR/R image for WTe\textsubscript{2} for probe polarization \protect\begin{math}\mathbf{\eta} || \mathbf{y}\protect\end{math} taken at \protect\begin{math}\sim\protect\end{math}710 $\upmu$J/cm\textsuperscript{2} absorbed pump fluence and T=80 K. (b) Fast Fourier Transform images performed across a few picoseconds after time zero after subtracting from the data a purely exponential fit based on Eq. (1) in the main paper for various probe photon energies.}
 \label{fig:80Kbhigh}
 \end{figure*}

\begin{figure*}[h!]
 \includegraphics[width=\textwidth]{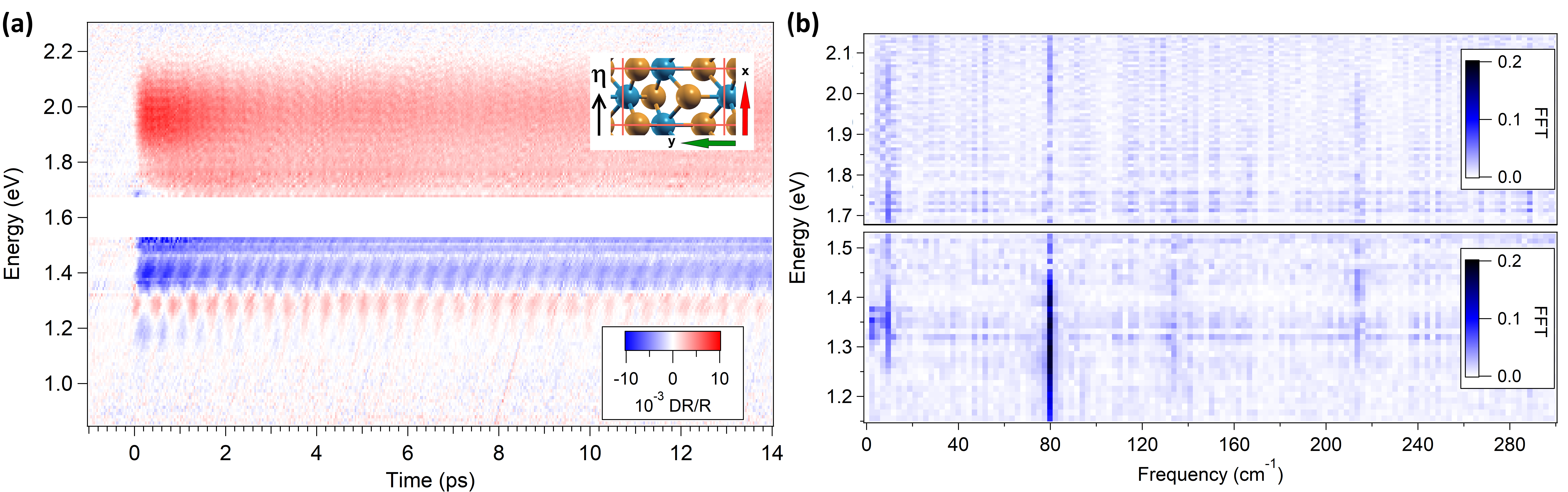}
 \caption{(a) Ultrafast broadband DR/R image for WTe\textsubscript{2} for probe polarization \protect\begin{math}\mathbf{\eta} || \mathbf{x}\protect\end{math} taken at \protect\begin{math}\sim\protect\end{math}230 $\upmu$J/cm\textsuperscript{2} absorbed pump fluence and T=80 K. (b) Fast Fourier Transform images performed across a few picoseconds after time zero after subtracting from the data a purely exponential fit based on Eq. (1) in the main paper for various probe photon energies.}
 \label{fig:80Kalow}
 \end{figure*}

\begin{figure*}[h!]
 \includegraphics[width=\textwidth]{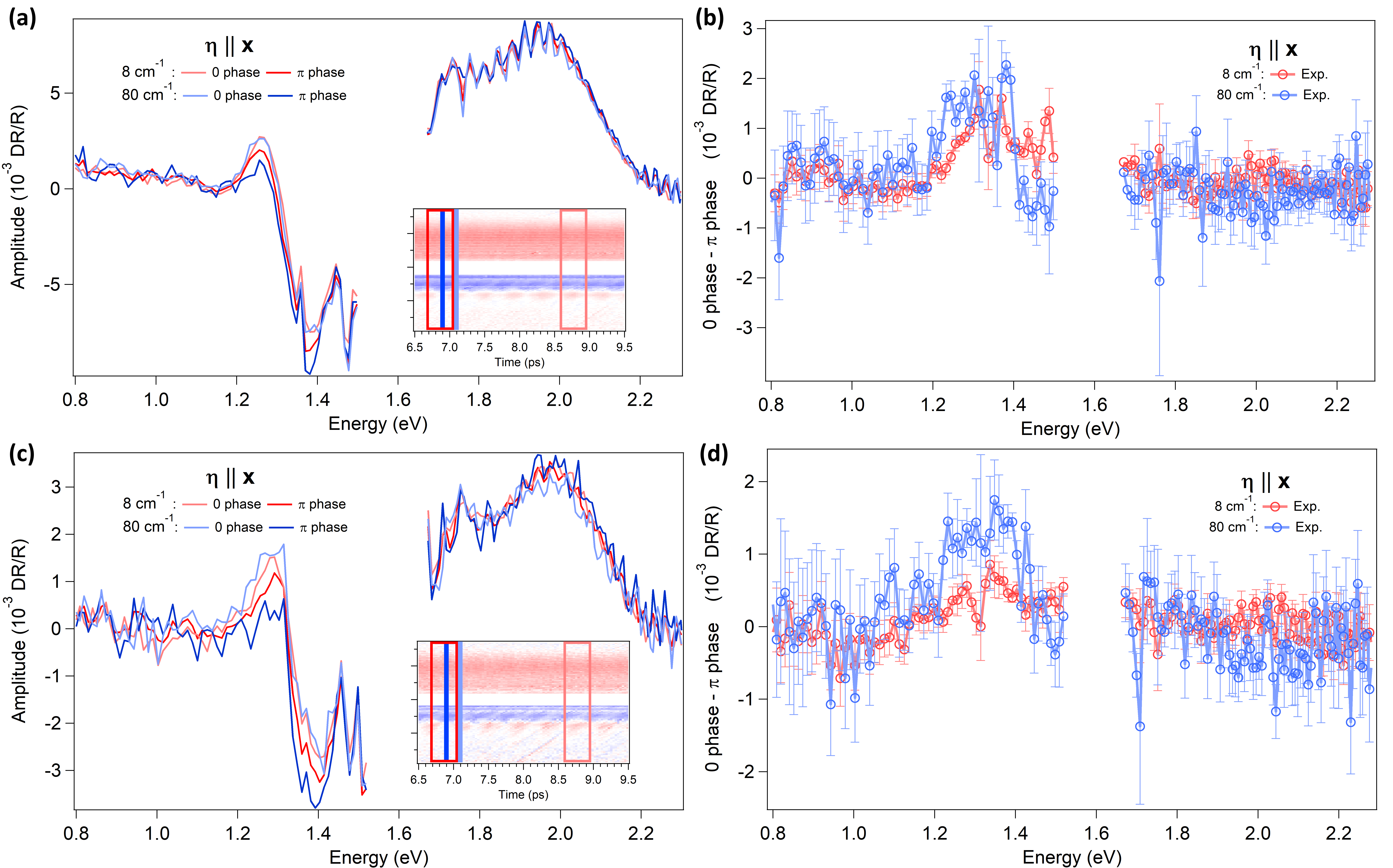}
 \caption{Spectral profiles taken at selected time delays for maxima and minima of the 8 cm\textsuperscript{-1} and 80 cm\textsuperscript{-1} optical phonon effects taken at a) \protect\begin{math}\sim\protect\end{math}710, c) \protect\begin{math}\sim\protect\end{math}230 $\upmu$J/cm\textsuperscript{2} absorbed pump fluence and T=80 K; the inset, with the same vertical axis and color scale as the corresponding graphs in Figs. \ref{fig:80Kahigh}, \ref{fig:80Kalow}, shows the exact temporal positions. (b),(d) Experimental DR/R difference between the 0 and \protect\begin{math}\pi\protect\end{math} phases displayed in respectively (a) and (c), showcasing the effects of the 8 cm\textsuperscript {-1} and 80 cm\textsuperscript {-1} optical phonons for \protect\begin{math}\mathbf{\eta} || \mathbf{x}\protect\end{math}.}
 \label{fig:80Kxmix}
 \end{figure*}

At T=80 K, phonons exhibit a blueshift in their frequencies with respect to T=295 K, as previously reported in \cite{slowph}\cite{Raman}. Compared to T=295 K the contribution from coherent phonons with frequency at higher than 80 cm\textsuperscript {-1} becomes more prominent in the DR/R images measured at \begin{math}\sim\end{math}710 $\upmu$J/cm\textsuperscript{2} (Figs. \ref{fig:80Kahigh}, \ref{fig:80Kbhigh}) and \begin{math}\sim\end{math}230 $\upmu$J/cm\textsuperscript{2} (Fig. \ref{fig:80Kalow}) absorbed pump fluence.  The phonon effects for \protect\begin{math}\mathbf{\eta} || \mathbf{x}\protect\end{math} shown in Fig. \ref{fig:80Kxmix} are analogous to the room temperature results reported in Figs. 4(a), \ref{fig:lowfluence}(d), although it becomes harder to separate them using the same method. This becomes especially challenging for  \protect\begin{math}\mathbf{\eta} || \mathbf{y}\protect\end{math}, where the amplitude of phonons with frequency higher than 80 cm\textsuperscript {-1} is comparable to the lower modes (Fig. \ref{fig:16supp}) throughout the investigated spectral region. In this case, a more advanced procedure involving a time-weighted average over multiple 0 or \begin{math}\pi\end{math} phase profiles to determine the initial amplitudes of the atomic shifts could be employed.

\subsection{2.4 - Ultrafast single-color probe optical spectroscopy}
\begin{figure*}[h!]
 \includegraphics[width=\textwidth]{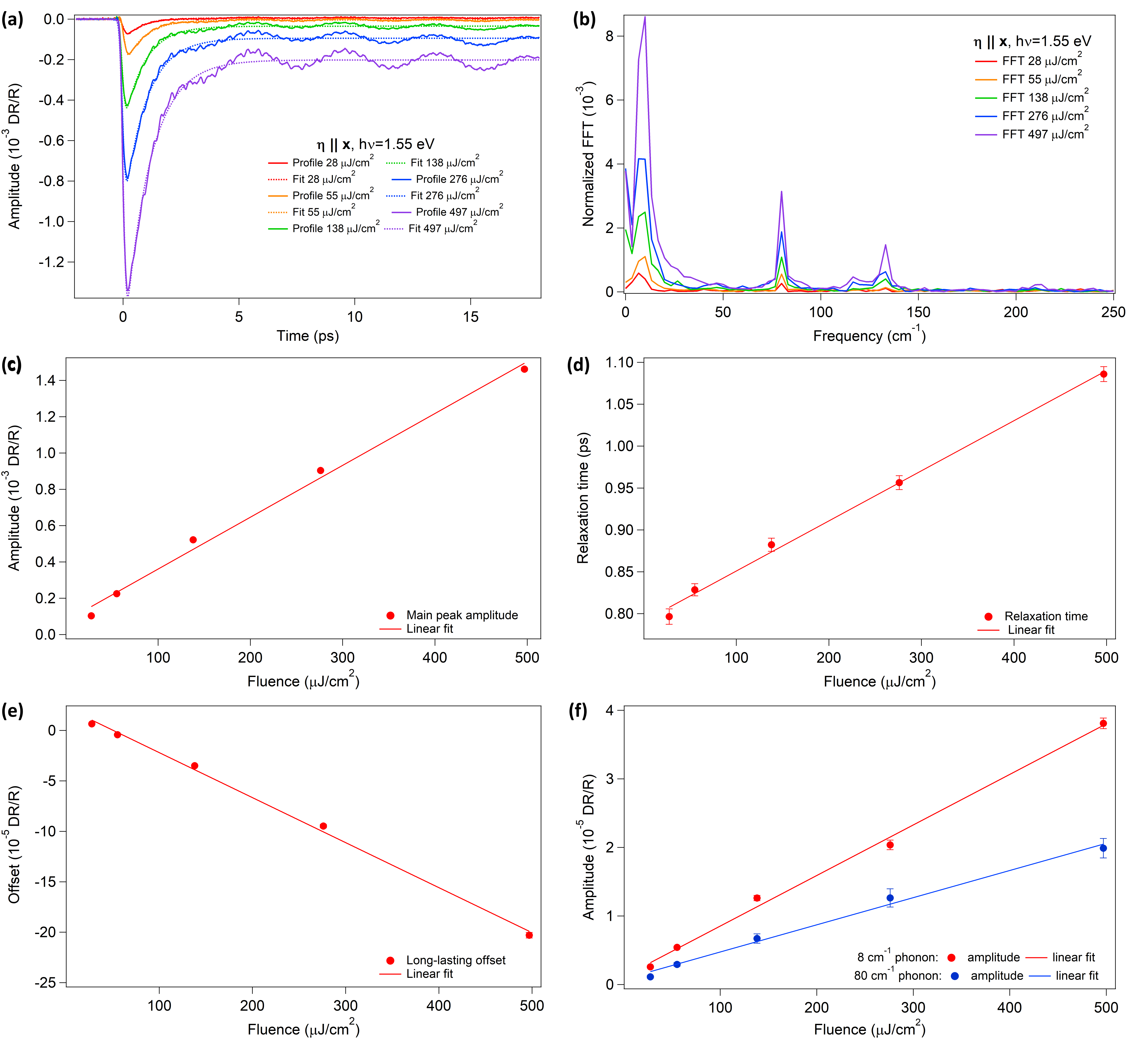}

\caption{(a) Absorbed fluence dependence on the out-of-equilibrium profiles performed with 1.03 eV pump and 1.55 eV single-color probe with \protect\begin{math}\mathbf{\eta} ||\mathbf{x}\protect\end{math}, at T=295 K, where a single-exponential decay model was used to fit the curves; see Eq. (1) of the main paper for details. (b) Fast Fourier Transform (FFT) transform of the residual signals after subtracting to the profiles the fit function. Fit parameters of the single-exponential decay model obtained from the curves reported in panel (a) respectively for (c) amplitude and (d) relaxation time of the main peak, (e) offset and (f) amplitudes of the phonon effects related to the 8 cm\textsuperscript{-1} and 80 cm\textsuperscript{-1} modes. These last values represent the amplitude of two damped cosine waves fitted to the residual signals.}

 \label{fig:295fluence}
 \end{figure*}

To investigate the response of the material as a function of the fluence, we performed a series of pump-probe reflectivity experiments at 1.03 eV pump and 1.55 eV probe photon energies. The results are summarized in Fig. \ref{fig:295fluence}. From Figs. \ref{fig:295fluence}(c),(f), it can be inferred that the amplitude of the main peak and of the phonon effects can be considered a linear function of the absorbed pump fluence in the whole investigated range within a very good approximation. Compared to the time-resolved broadband optical experiments, the 80 cm\textsuperscript{-1} oscillations are quenched because of the following effect. The probe pulse (1.55 eV, 800 nm) produced by the amplifier has a wide bandwidth (\textgreater 40 nm) with a finite residual chirp. This chirp cannot be resolved when the full bandwidth of the pulse is measured with a single photodiode, as it happens with lock-in acquisition, leading to a smearing of the fast oscillatory signal.
We also want to stress the fact that only the DR/R as measured by broadband spectroscopy can be considered as quantitative. The DR/R magnitude measured through lock-in acquisition is smaller by a given factor, arising from the signal treatment introduced by the lock-in device.

\begin{figure*}[h!]
 \includegraphics[width=\textwidth]{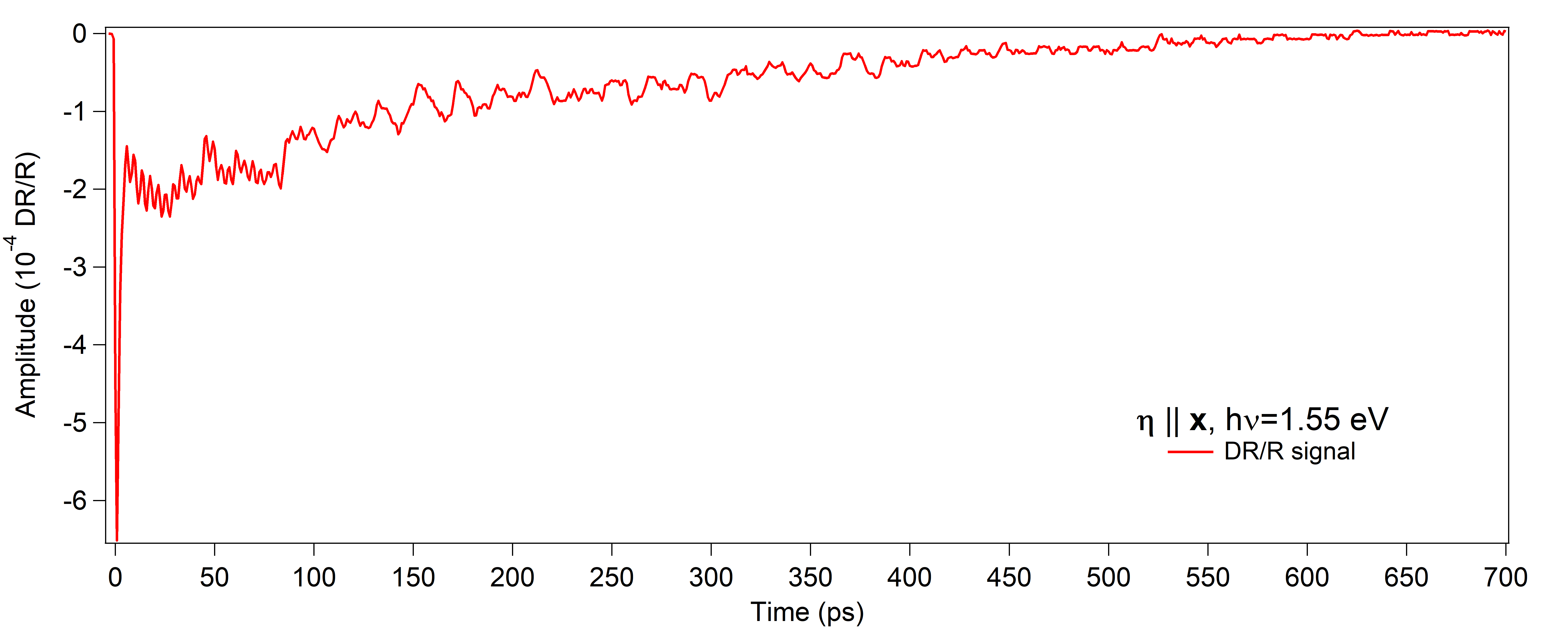}

\caption{Single-color DR/R curve obtained at 1.03 eV pump photon energy and a perpendicularly polarized 1.55 eV  probe beam polarized along \begin{math}\mathbf{\eta} || \mathbf{x}\end{math} at \begin{math}\sim\end{math}500 $\upmu$J/cm\textsuperscript{2} absorbed pump fluence and T=295 K, showing that the material returns to its equilibrium in about 700 ps.}
 \label{fig:295verylong}
 \end{figure*}

Furthermore, it is possible to determine the timescale of the long-lasting offset observed in the time-resolved profiles. As shown in Fig. \ref{fig:295verylong}, the signal survives for hundreds of picoseconds. Its timescale reminds of similar long-lasting non-equilibrium signal observed in time-resolved x-ray diffraction and time-resolved electron diffraction for transition metals, ascribed to the transient lattice heating \cite{latticeheating}.
 
\clearpage

\section{Part 3 - Computational results}
Density functional theory (DFT) simulations were carried out using norm-conserving (NC) \cite{ncoriginal} scalar relativistic \cite{Takeda1978} pseudopotentials with the generalized gradient approximation (GGA) in the Perdew-Burke-Ernzerhof (PBE) parametrization for the exchange-correlation functional \cite{pbe} chosen from the PseudoDojo database \cite{onvcsg15}\cite{pseudodojo}.  An orthorhombic simulation cell (Figs. \ref{fig:projection}(a)-(c)), with lattice constants a=3.477 {\AA}, b=6.249 {\AA} and c=14.018 {\AA} along the \textbf{x}, \textbf{y} and \textbf{z} directions respectively taken from the crystallographic data in \cite{cryst}, was used. It contains four tungsten and eight tellurium  atoms, with 14 and 16 valence electrons respectively, for a total number of 184 electrons per unit cell.

Structural optimizations and phonon calculations in the framework of density functional perturbation theory (DFPT) were performed using the \textsc{Quantum Espresso} (QE) \cite{QE-2009} suite of codes with a plane wave  kinetic energy cutoff of 70 Rydberg and a 12$\times$10$\times$6  uniform k-point mesh for integrations over the Brillouin zone.
Van der Waals contributions were not included, since, although in general relevant to describe layered compounds, in this specific case they do not improve the description of structural and electronic/optical properties of the system.

Td-WTe\textsubscript2 presents 33 zone-center optical phonon modes with four possible symmetry representations given by the C\textsubscript{2v} point group \cite{cryst}. In the following sections, we will focus on the two lowest energy A\textsubscript1 modes.
We calculated the diagonal macroscopic dielectric tensor components  through the \textsc{Yambo} code \cite{YAMBO} starting from the wavefunctions and eigenvalues obtained through the PW package of QE using the same kinetic energy cut-off and a 16$\times$14$\times$10 k-grid. 
Convergence tests showed that these parameters ensure the numerical accuracy needed to appreciate the relative effects of subpicometer atomic displacements on the electronic states in the near-infrared and visible spectral regions. The number of conduction bands to be considered in the calculations was also determined from convergence tests.
The reflectance curves were derived at the independent-particle (IP) level using a linear response approach.

\subsection{3.1 - Equilibrium reflectance}
\begin{figure*}[h!]
 \includegraphics[scale=0.40]{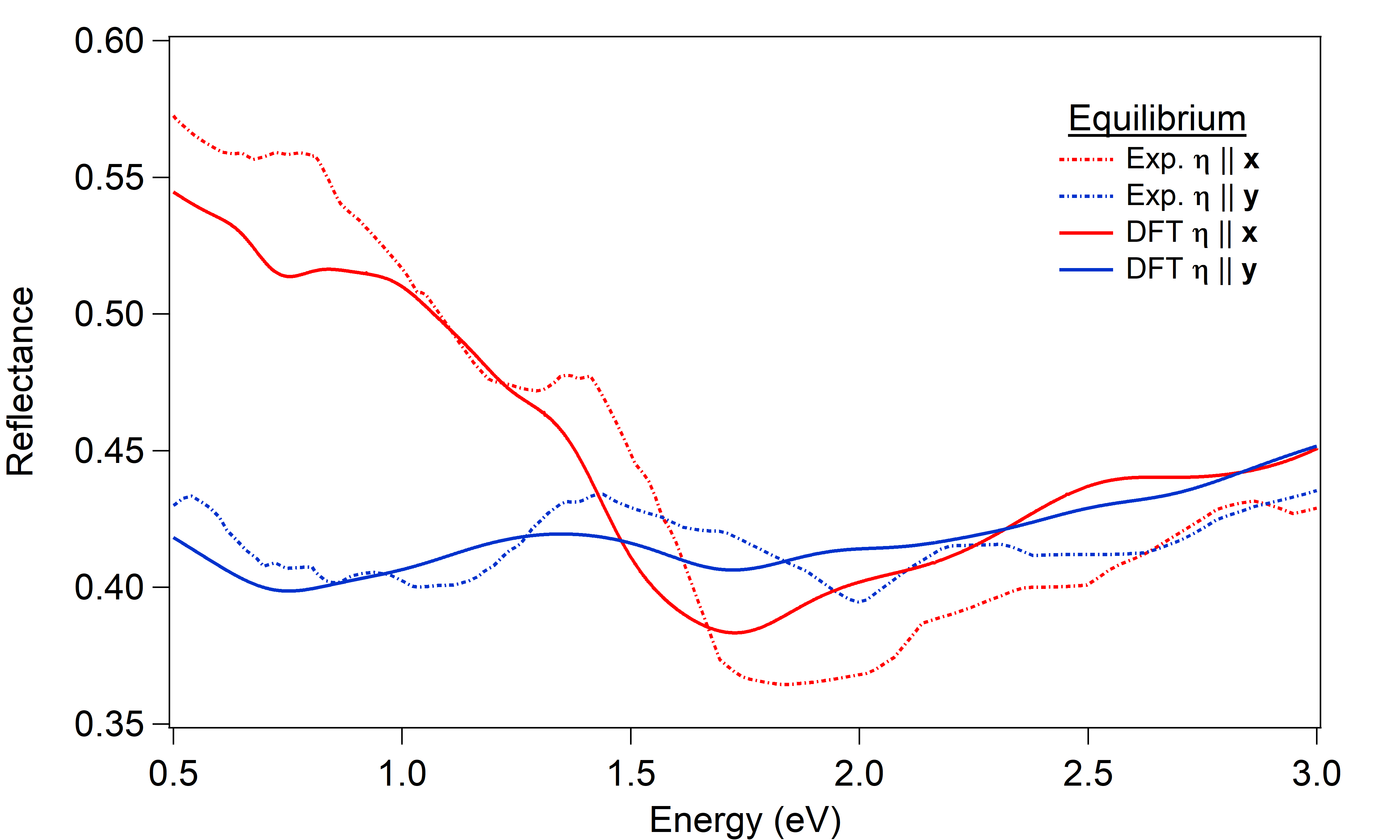}

\caption{Equilibrium reflectance curves in the two perpendicular probe polarizations; the experimental curves were derived from the data present in \cite{anieq} measured at 10 K.}

 \label{fig:dftvsexp}
 \end{figure*}

We report the calculated reflectance curves along the two perpendicular polarization directions in Fig. \ref{fig:dftvsexp}, compared with the reflectance data found in \cite{anieq}. We included 119 bands for the response function to reproduce the experimental data with a very good quantitative agreement. Although the experimental data was measured at 10 K and DFT results were obtained at T=0 K, they can be still considered adequate for our DR/R comparisons at T=295 K in good approximation, since the optical properties in this spectral region are only slightly affected by temperature \cite{eqrefl}.

\subsection{3.2 - Coherent optical phonon effects on the energy levels}
\begin{figure*}[t]
 \includegraphics[width=\textwidth]{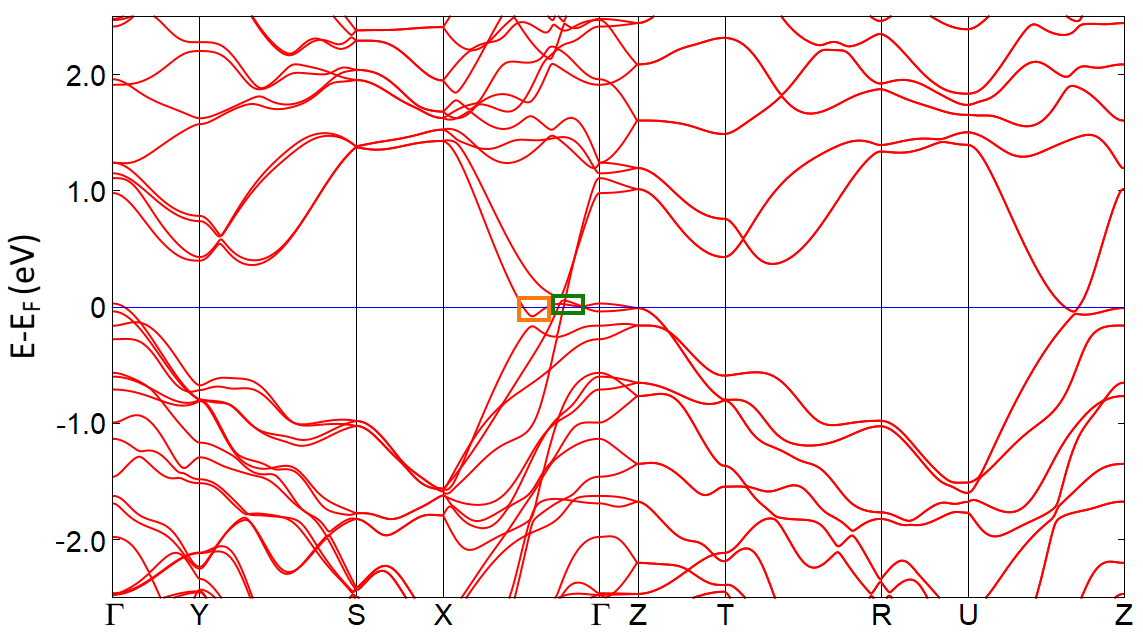}

\caption{Electronic band structure of WTe\textsubscript{2} in the equilibrium configuration; the orange and green rectangles highlight respectively the electron and hole pockets; the energy values are referred to the Fermi energy (E\textsubscript{F}).}

 \label{fig:BANDSeq}
 \end{figure*}

The subpicometer atomic displacements in the out-of-equilibrium structural configurations lead to modifications of the electronic band structure (Fig. \ref{fig:BANDSeq}) of the order of a few meV. 

\begin{figure*}[h!]
 \includegraphics[width=\textwidth]{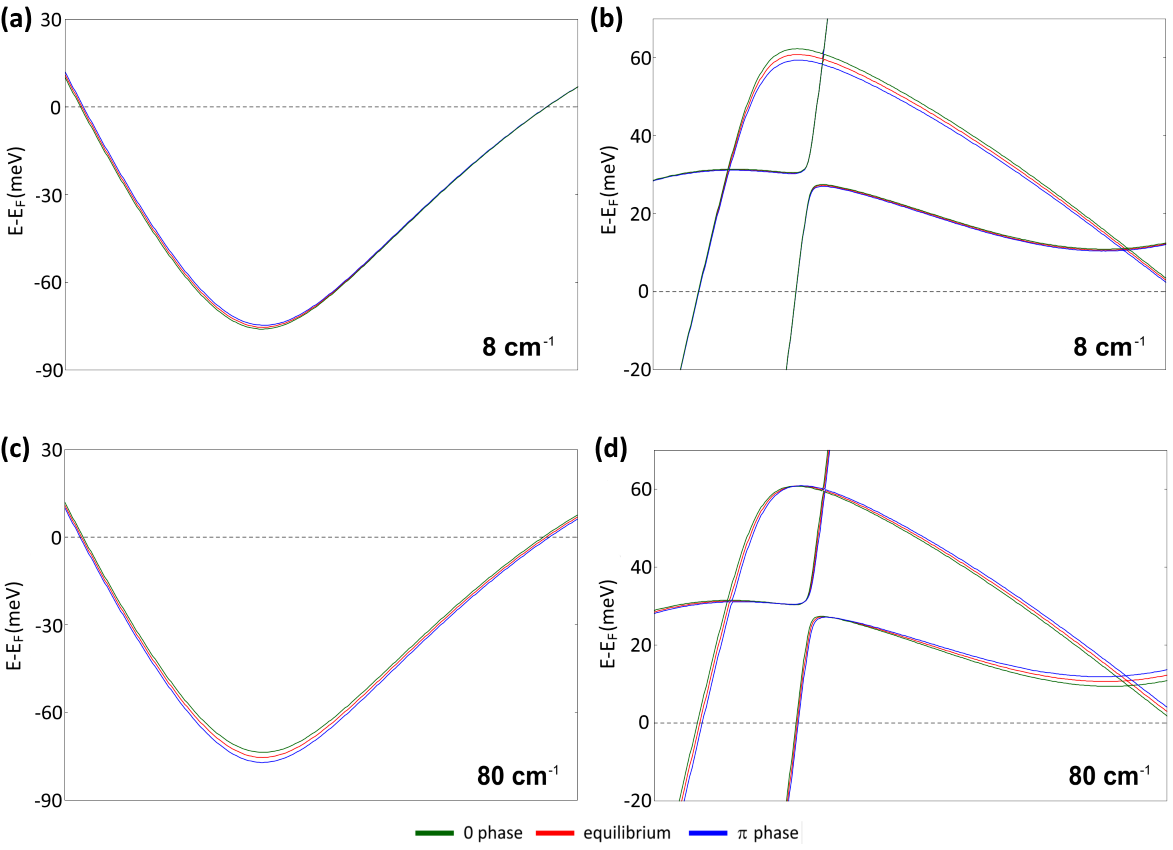}

\caption{
Effect of the phonon displacements on electron- and hole-like bands cut by the Fermi level along \protect\begin{math}\Gamma - X\protect\end{math}. (a),(b) Magnified views of the regions delimited by the orange and green rectangles in Fig. \ref{fig:BANDSeq} for the 8 cm\textsuperscript{-1} optical phonon mode; (c),(d) as (a),(b), but for the 80 cm\textsuperscript{-1} phonon mode; the 0 and \protect\begin{math}\pi\protect\end{math} phases are labeled as in the main paper, taking as reference the \protect\begin{math}\mathbf{\eta} || \mathbf{x}\protect\end{math} polarization; the displacements used for the graphs are double with respect to the low fluence values reported in the main paper to highlight the phonon effects.}

 \label{fig:BANDSwiggles}
 \end{figure*}

\begin{figure*}[h!]
 \includegraphics[scale=0.20]{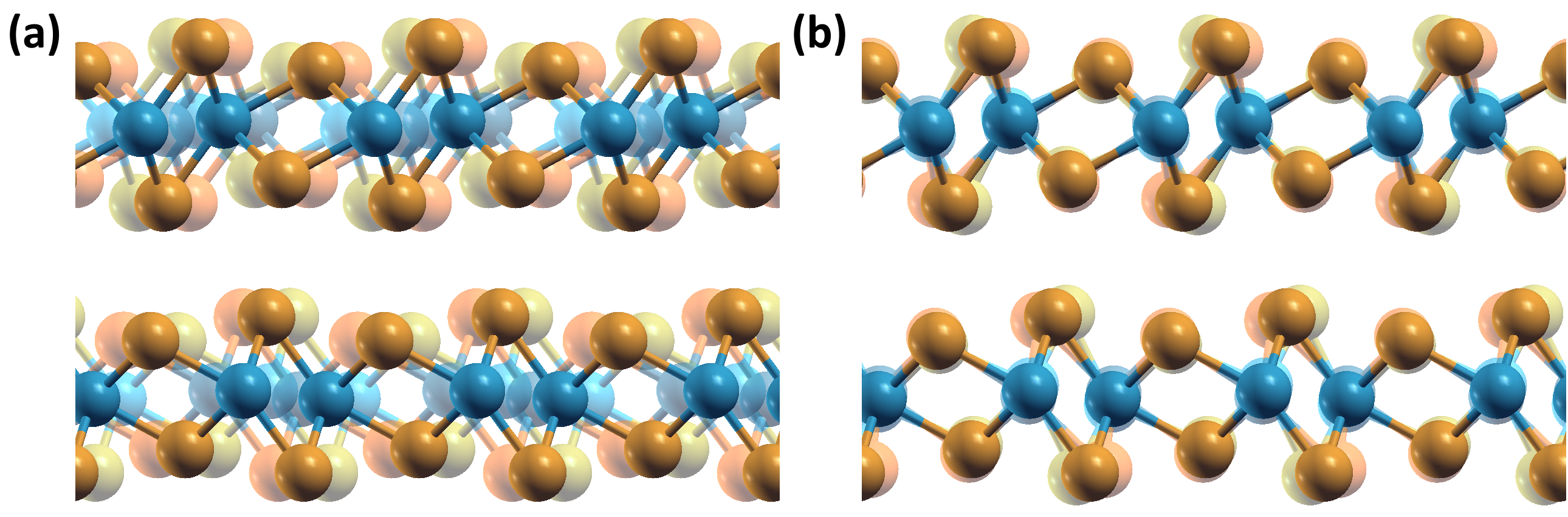}

\caption{The 0 and \begin{math}\pi\end{math} phases of the phonon oscillations for the (a) 8 cm\textsuperscript{-1} mode and (b) 80 cm\textsuperscript{-1} mode. The 0 phase is represented with transparent yellow Te atoms and light blue W atoms, while the \begin{math}\pi\end{math} phase is reported with transparent orange Te atoms and dark blue W atoms. Opaque colors describe the atoms in their equilibrium positions. The displacements are exaggerated for a better visualization.}

 \label{fig:0pi}
 \end{figure*}

Such variations are small, but beyond our relative numerical accuracy. Furthermore, we checked that these effects are almost linear with the displacements if we increase them by a factor up to twenty. We focus on the extrema of the electron and hole bands along \begin{math}\Gamma\end{math}-X. In Fig. \ref{fig:BANDSwiggles}, we show the results corresponding to the displacements estimated from the low (\begin{math}\sim\end{math}230 \textmu J/cm\textsuperscript{2}) fluence measurements and reported in the main paper. The effects (doubled in the graph for clarity) depend on the band and phonon mode. For the 8 cm\textsuperscript{-1} mode, the bottom (top) of the electron (hole) band at -80 meV (60 meV) under (above) the Fermi energy experiences a \begin{math}\sim\end{math}0.7 (1.5) meV difference between the two oscillation phases (0 and \begin{math}\pi\end{math}, represented in Fig. \ref{fig:0pi}), while for the 80 cm\textsuperscript{-1} modes the difference becomes \begin{math}\sim\end{math}1.7 (0) meV. By properly tuning the Fermi level, these variations, although small, could have an impact on the extremely high magnetoresistance in WTe\textsubscript{2} \cite{magnetoWTe2}, which is ascribed to a delicate balance between electron and hole carriers. Additional contributions may come from carrier mobility and the magnetic field \cite{Carriermobility} \cite{Magneticfield}.
We caution the reader that, as stated in the main paper, we used scalar relativistic pseudopotentials. The inclusion of spin-orbit coupling (SOC) using fully relativistic pseudopotentials would provide a more accurate description of these effects, since it is known that SOC plays an important role on the electronic bands around the Fermi level in WTe\textsubscript{2} \cite{weyl2}.

\subsection{3.3 - Bands contribution to the response function}
\begin{figure*}[h!]
 \includegraphics[width=\textwidth]{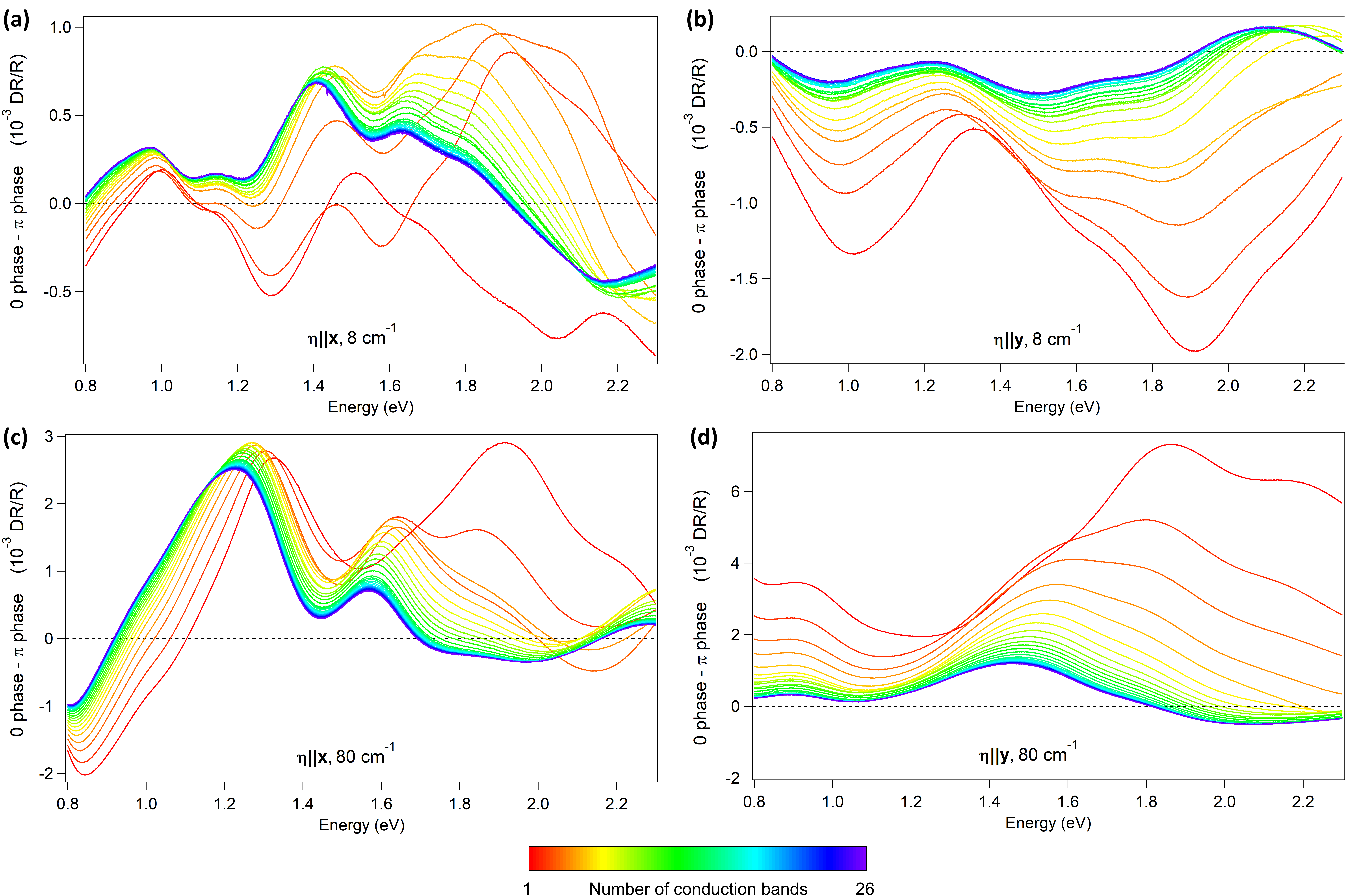}

\caption{Phonon effects on the DR/R profile obtained from the differences between the 0 and \begin{math}\pi\end{math} phases with an increasing number of conduction bands inserted in the response function; (a) 8 cm\textsuperscript{-1} \protect\begin{math}\mathbf{\eta} || \mathbf{x}\protect\end{math}; (b) 8 cm\textsuperscript{-1} \protect\begin{math}\mathbf{\eta} || \mathbf{y}\protect\end{math}; (c) 80 cm\textsuperscript{-1} \protect\begin{math}\mathbf{\eta} || \mathbf{x}\protect\end{math}; (d) 80 cm\textsuperscript{-1} \protect\begin{math}\mathbf{\eta} || \mathbf{y}\protect\end{math}.}

 \label{fig:bandsconv}
 \end{figure*}

To accurately describe the effects induced by the optical phonons, it is necessary to include in the response function an appropriate number of bands around the Fermi level. The ''converged'' curves shown in Fig. \ref{fig:bandsconv} were obtained including 119 bands, which is safe enough compared with the 184 electrons per unit cell and the spectral range under consideration for the transitions.  In the same figure, we show the effect of varying the number of ''conduction bands'', i.e. the totally unoccupied bands above the Fermi level. We notice that the main features of the DR/R profile are quickly captured even using few conduction bands, while to correctly compare the DFT amplitudes with the experimental data and give a reasonable estimate of the ionic displacements, more than 15 conduction bands have to be included.

However, since major features are still retained adding only the first conduction band in the response function, we deduce that transitions having final states that belong to bands crossing the Fermi level are involved. This supports the claim that for this material the phonon effects cannot be attributed to single transition in k-space, but to multiple contributions.
 
\clearpage
\bibliography{SMa}